\documentclass[11pt,preprint]{aastex}
\usepackage{lscape}
\usepackage{natbib}
\usepackage{chngpage}

\title{High Resolution H~I Distributions and Multi-Wavelength Analyses of Magellanic Spirals NGC 4618 and NGC 4625}
\author{Jane F. Kaczmarek\footnotemark[1]  \& Eric M. Wilcots}
\affil{University of Wisconsin Madison, Department of Astronomy, 475 N. Charter
Street, Madison WI 53706 USA}
\email{jane@astro.wisc.edu, ewilcots@astro.wisc.edu}

\begin{abstract}

We present a detailed analysis of high resolution H~I observations of
the Magellanic spiral galaxies NGC 4618 and NGC 4625.  While the H~I
disk of NGC 4625 is remarkably quiescent with a nearly uniform
velocity dispersion and no evidence of H~I holes, there is a dynamic
interplay between star formation and the distribution of neutral
hydrogen in NGC 4618.  We calculate the critical density for
widespread star formation in each galaxy and find that star formation
proceeds even where the surface density of the atomic gas is well
below the critical density necessary for global star formation.  There
are strong spatial correlations in NGC 4618 between UV emission, 1.4
GHz radio continuum emission, and peaks in the H~I column density.
Despite the apparent overlap of the outer disks of the two galaxies,
we find that they are kinematically distinct, indicating that NGC 4618
and NGC 4625 are not interacting.  The structure of NGC 4618 and, in
particular, the nature of its outer ring, are highly suggestive of an interaction, but the timing and nature of such an interaction remain unclear.

\end{abstract}

\begin{document}
\footnotetext[1]{\footnotesize Now at Sydney Institute for Astronomy, School of Physics H90, The University of Sydney, NSW 2006, Australia}
\maketitle

\section{Introduction}

One of the more stunning revelations about the Local Group in the past
few years has been the observations of the proper motions of the
Magellanic Clouds showing that they are moving at significantly higher
velocities than previously thought (e.g. Kallivayalil et al. 2006,
Piatek et al. 2008, Besla et al. 2010).  One major implication of
these results is that the Magellanic Clouds may very well be on their
first passage through the Local Group, contradicting the long-held
view that they are long-standing companions to the Milky Way.  From
another point of view, this argument makes sense.  The Magellanic
Clouds, and the Large Magellanic Cloud in particular, are unlike any
other companions of the Milky Way, M31, or the vast majority of other
spiral galaxies (e.g. James \& Ivory 2011).  In addition, over the
past decade or so we have seen a proliferation of statistically
significant samples of galaxies that show that objects sharing the
basic morphology of Magellanic spirals are common in
both the local Universe and at intermediate redshift (Garland et
al. 2004, Ryan-Weber et al. 2003, Sheth et al. 2008).  Part of the 
motivation for the study described here is that the
dynamics, structure, and star formation history of the LMC have long
been interpreted in the context of its proximity to both the Milky
Way and the SMC.  The fact that there are so many galaxies that share
the LMC's structure but not its proximity to other massive galaxies,
and the possibility that the LMC has only recently entered the Local
Group drives us to examine the properties of the larger population of
Magellanic spirals in detail sufficient to meaningfully and properly 
understand how the LMC fits into the larger context of late-type
barred galaxies.

The earliest comprehensive look at Magellanic spirals was carried out
by de Vaucouleurs \& Freeman (1972) who noted the structural
similarity between the LMC and a number of other nearby galaxies.
Much of the subsequent work was aimed at understanding the origin of
the lopsided structure that characterized Magellanic type galaxies.
Interactions were primarily believed to be responsible for
the asymmetric properties of this class of galaxy; a scenario which
was well fit by the LMC itself.  Theoretical work and some simulations
could also accurately account for the apparent lopsidedness of
Magellanic spirals, but not their frequency.

Baldwin et al. (1980) originally suggested that differential
precession of the disk in an axisymmetric potential would give rise to
the asymmetry characteristic of SBm galaxies.  Their proposal,
however, predicted that the winding problem would result in the
asymmetry dispersing over 5 Gyr, too short a timescale to account for
the ubiquity of lopsidedness.  Walker, Mihos, \& Hernquist (1996) and
Zaritsky \& Rix (1997) proposed that minor mergers could result in
lopsided distributions.  Compelling evidence for the role of
interactions in accounting for lopsided morphology comes from a
statistical survey based on examination of POSS and UK Schmidt plates
that found 71/75 Magellanic spirals with well-defined one armed
morphologies appear to have a physical neighbor (Odewahn 1994).
Wilcots \& Prescott (2004), however, showed that only 4 of 13 galaxies from the
Odewahn (1994) sample were actually interacting and Wilcots, Lehman
\& Miller (1996) showed that even those that were interacting did so
weakly.  These studies weakened the argument that asymmetry is
generally caused by interactions, and bolstered a model put forth by
Levine and collaborators (Levine \& Sparke 1998, Noordermeer, Sparke,
\& Levine 2001) that lopsidedness is stable once the disk is displaced
from the dynamical center of the halo and is allowed to rotate around
the center.

One example of an instance in which an interaction may be linked to
the characteristics of a Magellanic spiral is the NGC 4618-4625 pair
of galaxies.  NGC 4618 and NGC 4625 have both been classified as
Magellanic-type spirals (Odewahn 1994), though only NGC 4618 exhibits
the classic optical morphology of a barred galaxy in this class (de
Vaucouleurs \& Freeman 1972).  Tully (1988) gives a distance of 7.3
Mpc to NGC 4618 and 8.2 Mpc NGC 4625, but their apparently
overlapping H~I distributions lends support to the notion that
interactions play a role in generating the characteristic morphology
(Bush \& Wilcots 2004). In addition, recent {\it GALEX} observations reveal NGC 4625 to
have an extended UV disk as well as multiple branching, ragged spiral
arms.  NGC 4625 has been noted to have one of the largest UV to
optical disk (as measured by $D_{25}$) ratios of 4 (Gil de Paz et
al. 2005) and an even larger H~I to optical disk ratio of 9.8 (Bush \&
Wilcots 2004).  It is remarkably symmetric and Bush \& Wilcots (2004)
noted that it seems unaffected by its apparent interaction with NGC
4618.  From its morphology alone, NGC 4618 appears to be a
prototypical Magellanic spiral, with an obvious stellar bar and a
lopsided one-armed spiral structure.  Both galaxies are host to
on-going star formation; in NGC 4618 most of the H~I regions lie near
the central bar, while in NGC 4625 the star formation is nearly
uniformly distributed across its disk.

The goals of this work are to re-examine the issue of whether these
two galaxies are indeed interacting and what effect such an interaction
may be having on both NGC 4618 and NGC 4625, particularly with regard
to star formation.  We also seek to better understand the symbiotic 
relationship between the neutral atomic interstellar medium and star 
formation in Magellanic spirals.  In this paper
we present a detailed analysis of high-resolution observations of the
neutral gas content of NGC 4618 and NGC 4625, complementing the
earlier work of Bush \& Wilcots (2004) (hereafter, BW04).  We describe
our observations in $\S$2 and present the H~I distribution and overall
kinematics in $\S$3.1. We show the tilted ring models of rotation
curves for each galaxy in $\S$3.2. We begin our analysis in $\S$4 by looking at
the distribution of massive stars and the structure and kinematics of the
neutral gas in each galaxy. $\S$4.1
identifies the H~I holes in the center of NGC 4618 and $\S$4.2 looks
into correlations with tracers of massive star formation. $\S$4.3 calculates star
formation thresholds and the corresponding measured column
densities. The velocity dispersion of each H~I disk is addressed in
$\S$4.4. We discuss the possibility of an ongoing interaction between
NGC 4618 and NGC 4625 in $\S$5 and finish our analysis in $\S$6 where
we look into the nature of the H~I ring around NGC 4618. Our
conclusions are stated in $\S$7.
\footnotetext{\footnotesize The National Radio Astronomy Observatory is a facility of the National
Science Foundation operated under cooperative agreement by Associated
Universities, Inc.  This work also made use of observations made with
the NASA Galaxy Evolution Explorer.  {\it GALEX} is operated for NASA by the
California Institute of Technology under NASA contract NAS5-98034.}

\section{Observations and Data Reduction}

We observed the NGC 4618/4625 system using the Very Large Array in the
B and C configurations for total integration times of 36 hours and 20
hours, respectively. After each 50 minute exposure series, a 3 minute
observation was taken of the phase calibrator, 1225+368. Two separate
observations were made of the amplitude and bandpass calibrator,
1328+307, for 10 minutes each.  The AIPS tasks \textit{`vlacalib'} and
\textit{`vlaclcal'} were employed to calibrate both the amplitude and
phases, respectively. In order to subtract the continuum and clean our
data further, several line-free channels were fit and then subtracted
off using the AIPS task \textit{`uvlin'}.  The task \textit{`imagr'}
was used to create and clean both a naturally weighted (robust 5) and
intermediately weighted (robust 0) image cube of the galaxy
systems. 1000 clean iterations were done to remove the side lobes from
our intermediately weighted cube and 1800 for our naturally weighted
cube. Our beam size was 13.14''$\times$ 13.02'' for the naturally
weighted and 10.02''$\times$9.40'' for the intermediately weighted
cubes. We compare this to the resolution of BW04 of 19.6" $\times$
16.9". The angular extent of 1 arcsecond corresponds to 35.4 pc for the distance to NGC 4618 (7.3 Mpc) and 39.7 pc for the distance to NGC 4625 (8.2 Mpc) on the
sky. Our cube has a velocity resolution of 5.2 km s$^{-1}$ and spans a
velocity range of 255.9 -- 850.4 km s$^{-1}$. We adopt
redshift-independent distances in our analysis of the galaxies in
order to avoid any local gravitational effects on the measured
distances. From Tully (1988), we use a distance of 7.3 Mpc and 8.2 Mpc
to NGC 4618 and NGC 4625, respectively.

\section{Results}

\subsection{Distribution and Kinematics of H~I}

A naturally weighted cube was used to produce Figures 1-5. A
flux cut off of 2$\sigma$ (0.64$\times$10$^{-4}$ Jy B$^{-1}$ or
7.3$\times$10$^{-1}$ M$_{\odot}$pc$^{-2}$) was applied to the resulting
images: the integrated-flux density (moment 0), the velocity field
(moment 1), and the velocity dispersion (moment 2). We also create an
intermediately weighted data cube (robust 0) and apply a flux cut off
of 2$\times$10$^{-4}$ Jy B$^{-1}$ (2.3$\times$10$^{-1}$
M$_{\odot}$pc$^{-2}$). We do this with the motivation of taking
advantage of the higher resolution data available for analysis of the
H~I holes, star-formation thresholds and investigation of the overall
fine structure of the galaxies. 

Figure 1 shows the emission of each channel of the naturally weighted
cube and Figure 2 shows the integrated-intensity map of NGC 4618 and
NGC 4625. NGC 4618 is located at (R.A. = 12$^{h}$41$^{m}$35$^{s}$,
dec = 41$^{\circ}$08'23''[J2000]) corresponding to the southwestern
portion of the image and NGC 4625 in the northeast corner is located
at (R.A. = 12$^{h}$41$^{m}$52$^{s}$, dec = 41$^{\circ}$16'18''
[J2000]). NGC 4625 first appears in the upper left panel of Figure 1
at a velocity of 648.8 km s$^{-1}$ with the initial emission from NGC
4618 appearing at a velocity of 617.7 km s$^{-1}$.  NGC 4625 disappears
by a velocity of 560.8 km s$^{-1}$.  The channel maps reveal the complex
extended structure of NGC 4618 which we characterize as a ring.  This
extended emission is notable as the gas north of the main body of NGC 4618
most readily seen in the channels displayed in the upper right panel
of Figure 1.  By the end of the cube in the lower left panel of Figure 1,
the extended gas is seen largely to the south of the main body of NGC 4618.

The large, extended H~I ring surrounding the disk of NGC 4618, as well
as a number of holes in the H~I distribution in the central regions of
the galaxy, are evident in the moment 0 map in Figures 2 and 3. Figure 3 represents the integrated H~I map overlaid on the SDSS r-band image of the NGC 4618 \& NGC 4625 field. The H~I contours and beam size are the same as in Figure 2. We identify
and discuss the holes in further detail in $\S$4.1 and the ring in
$\S$6.  Even at this higher resolution NGC 4625 seems to lack the H~I
holes we see in NGC 4618.  NGC 4625 also lacks the very high column
density peaks we see in NGC 4618.  From the integrated H~I column
densities, we measure average column densities of 9.3 and 3.9
M$_{\odot}$pc$^{-2}$ for the central, optically bright regions of NGC 4618 and NGC 4625,
respectively. An average density of 0.9 M$_{\odot}$pc$^{-2}$ is measured in
the ring of NGC 4618 and 3.02 M$_{\odot}$pc$^{-2}$ in the brightest
star-forming arm of NGC 4625 as identified in Figure 3.

Figure 4 shows the velocity field of both galaxies.  NGC 4625 shows little evidence of 
any perturbation or deviation
from a typical differentially rotating disk, though there may be
evidence of streaming caused by the spiral arm in the southwest corner
of the galaxy as indicated by kinks in the isovelocity contours
associated with the spiral arms.  The velocity field of NGC 4618,
however, is quite different. The velocity field of the gas that is
spatially coincident with the main stellar body shows a severe twist,
consistent with the ``S'' shaped isovelocity contours one expects to
see in a barred galaxy (Kalnajs 1978).  As BW04 showed, the outer ring
of the galaxy appears to be a separate kinematical component and not
necessarily a smooth extension of the inner disk.  We discuss the
origin and nature of the H~I ring later in $\S$6.

Finally, we show the velocity dispersion maps of each galaxy in Figure
5.  Once again, NGC 4625 appears to have a quiescent disk with nearly
constant velocity dispersion across its full extent.  We see the
velocity dispersion in both NGC 4618 and NGC 4625 peak in those same
regions where the H~I column density is highest.  The ring around NGC
4618 is also kinematically cold with an average velocity dispersion of
$\sim10$ km s$^{-1}$.  The most interesting feature in Figure 5 is the
very high values in the region where the disks of the two galaxies
appear to overlap.  We discuss in $\S$5 that this is simply a
reflection of the fact that the disks are kinematically distinct.

\subsection{Rotation Curves}

To analyze the nature of the kinematics in both galaxies in more detail we
attempted to fit tilted ring models to the velocity fields in Figure 4
using the AIPS task \textit{`gal'}.  In our model of NGC 4618, the
disk and ring of the galaxy are fit separately in order to accurately
map the entire observed extent of the galaxy. Attempts to fit the
galaxy as a continuous body resulted in nonsensical curves, likely due
to the large gap between the disk and the H~I ring. Throughout, we
allow the inclination and position angle to vary while keeping the
center and systemic velocity fixed. The position angle of the major
axis and the inclination of the disk are plotted with along with the
rotation curves of the corresponding section of the galaxy in Figure
6.

Our fit to the velocity field of NGC 4618 shows that the position
angle is nearly constant through the inner part of NGC 4618 out to a
radius of 55$^{\prime\prime}$. The sharp change in the inclination and rotational velocity at 50 - 55$^{\prime\prime}$ indicate the extent of the
stellar bar in the center of the galaxy.  Beyond a radius of 55$^{\prime\prime}$ the position angle smoothly changes by
~50$^{\circ}$ over 80$^{\prime\prime}$. The inclination smoothly
decreases from 80$^{\circ}$ to just below 30$^{\circ}$ over the inner
60$^{\prime\prime}$ and remains relatively constant beyond that.

To best represent NGC 4625, we average six different solutions of the
task \textit{`gal'}, simultaneously plotted below the final solution
in Figure 7. For each solution, the only variation between the input
parameters was slightly different position angles or inclinations,
each leading to a unique solution. The spiral arms of NGC 4625 are located
at a radius of 80 arcseconds which we see corresponds to the peaks in
the inclination, so it is likely that this feature is a result of 
streaming motions along the arms.

In both models of the galaxies, we see rotational velocities
consistent with other Magellanic and late-type spirals. Our results
are in agreement with those shown in BW04 and are comparable to van
Moorsel's (1983) previous work on NGC 4618 and NGC 4625. Our
differences likely stem from our better resolution. Our range of
inclination in the disk of NGC 4618 agree well with those in BW04;
however, we derive a narrower range of inclination values for the
outer ring which varies from 30$^{\circ}$ -- 45$^{\circ}$ and has a
trend of increasing inclination with increasing radius. Our results
for NGC 4625 also agree well with the previous rotation curves from
BW04, with our inclination having a smaller range than that of BW04
where the fitted inclination varied between 20$^{\circ}$ and
40$^{\circ}$ whereas our results span angles of 24$^{\circ}$ to
36$^{\circ}$.

\section{Massive Stars and the Interstellar Medium in NGC 4618 and NGC 4625}

Much of the motivation for this work was to better understand the
symbiotic relationship between massive stars and the distribution and
kinematics of the neutral interstellar medium in both galaxies.  We
approach this problem by looking at the properties of H~I holes in
both galaxies and by correlating observational tracers of massive star
formation with the distribution and kinematics of the H~I.  As nicely
summarized by Bagetakos et al. (2011), the issue of how feedback from
massive stars influences the interstellar medium has been studied
extensively for quite some time.  In the standard analysis, perhaps
best described by McCray \& Kafatos (1987) and Tenorio-Tagle et
al. (2005), H~I holes are the result of feedback from the evolution of
massive stars where powerful supernovae eventually clear out areas in
the interstellar medium, creating a hot bubble that will expand until
the velocity of the gas slows to that of the velocity dispersion of
the galaxy.  This model has formed the basis of the interpretation of
a number of studies of the distribution and properties of H~I holes in
nearby galaxies (e.g. Puche et al. 1992, Wilcots \& Miller 1998).
More recent analyses (e.g. Silich et al. 2006, Warren et al. 2011,
Weisz et all. 2009, Cannon et al. 2011) suggest a more complex
process, which involves multiple generations of star formation.  In an
analysis of H~I holes in galaxies in the THINGS survey Bagetakos et
al. (2011) conclude that holes are ultimately the result of feedback
from star formation.

In either scenario, the gas removed by the supernova bubble can
pile up in areas surrounding the gas-deficient holes where the
resulting increase in density can lead to positive feedback in the
form of increased star formation. In principle; one might therefore
expect the H~I holes to be expanding, contain a concentration of young
stars and be surrounded by ongoing star formation. A detailed
examination of the stellar content within the holes is beyond the
scope of this paper.

To further probe the nature of the star formation and feedback in both
galaxies we constructed maps of the 1.4 GHz radio continuum emission
associated with NGC 4618 and NGC 4625 by combining all the line-free
channels of our intermediate data cube to create a composite
image. The 1.4 GHz radio continuum traces both the thermal (free-free)
and non-thermal (synchrotron) emission associated with massive star
formation (Condon 1992).  The former is directly correlated with H~II
regions while the synchrotron emission is associated with supernovae
explosions (e.g. Chomiuk \& Wilcots 2009).  In either case the 1.4 GHz
radio continuum is an excellent tracer of massive stars. Having only a
single frequency to work with we cannot accurately separate the two
components, however Heesen et al (2011) find that 50\% of the 1.4 GHz
emission in the irregular galaxy IC 10 is thermal while Kepley et
al. (2011) found that up to 40\% of the 1.4 GHz emission in NGC 4214
is thermal. For our analyses, we take 50\% of our continuum emission
to be thermal. 

\subsection{Identification of H~I Holes}

H~I holes in the disk of NGC 4618 were first noted in BW04.  Utilizing
our new high resolution column density and velocity maps we identify
six distinct holes in NGC 4618.  The positions of the holes are listed
in Table 1 and shown in Figure 8.  All holes are defined as coherent
structures having a H~I surface density less than 6.42
M$_{\odot}$pc$^{-2}$.

\begin{table}[ht]
\centering
\caption{Properties of H~I Holes}
\centering
\begin{tabular}{|c c c|}
\hline\hline
Classified Hole Number & Center RA & Center Dec \\
\hline
1 & 12$^{h}$41$^{m}$30.9$^{s}$ & 41$^{\circ}$07'45''.8 \\
2 & 12$^{h}$41$^{m}$32.8$^{s}$ & 41$^{\circ}$07'54''.4 \\
3 & 12$^{h}$41$^{m}$35.5$^{s}$ & 41$^{\circ}$08'50''.3 \\
4 & 12$^{h}$41$^{m}$35.4$^{s}$ & 41$^{\circ}$10'07''.6 \\
5 & 12$^{h}$41$^{m}$32.4$^{s}$ & 41$^{\circ}$09'16''.0 \\ 
6 & 12$^{h}$41$^{m}$32.4$^{s}$ & 41$^{\circ}$08'50''.2 \\
\hline
\end{tabular}
\end{table}

To better understand the physical nature of the holes we examined the
kinematics of the H~I surrounding each one.  The velocity profiles
were plotted using the AIPS task \textit{plcub} where each
box represents the H~I profile (intensity vs. velocity) for a region
equivalent to $\sim250$pc on a side in the galaxy (Figure 9). The general absence of double
peaked and broad profiles of the surrounding gas makes is a compelling
argument that the H~I holes are not expanding.  More significantly,
there is essentially no velocity gradient across the hole as one would
expect if the holes were expanding.  Therefore, if the holes were
caused by stellar feedback, the events occurred long enough ago that
the expansion has stopped.  Without a detailed analysis of the
underlying stellar population we cannot put a constraint on the
timescale of such an event.

\subsection{Correlations with UV and 1.4 GHz Emission}

Even in the absence of analysis of the stellar population we can
correlate the position of the H~I holes with recent star formation as
traced by its UV emission as seen by {\it GALEX}.  In Figure 10 we plot
the NUV emission from {\it GALEX} with red
contours, the H~I column density in blue, and the 1.4 GHz continuum
emission (discussed in more detail below) in green.  Each is
overplotted on an image of NGC 4618 taken from a r-band Sloan Digital
Sky Survey DR7 (Abazajian et al. 2009).  The most obvious feature is
the apparent stellar bar in which we seen copious UV emission (red
contours) and enhanced 1.4 GHz emission.  There is an overall deficit
of H~I in the bar, similar to the lack of H~I seen in the stellar bar
of the LMC (Staveley-Smith et al. 2003).  While there is a significant
concentration of UV and 1.4 GHz emission associated with the bar in
NGC 4618, we also see an enhancement in the UV emission along the rim
of the large complex of H~I holes 1 \& 2 (Figure 8) in the southern
half of the galaxy.  The local peak of the UV emission in the arm
($12^{h}41^{m}33^{s}$, $+41^{\circ}08^{\prime}30^{\prime\prime}$) is
also coincident with a ridge of 1.4 GHz emission and an enhancement in
the underlying distribution of stars.  This type of phenomenology is
what one would expect if the expansion of the hole leads to an
enhancement in the ISM density and subsequent star formation along a
shell surrounding the hole.  Additional peaks in UV emission are in
close proximity to, but slightly offset from, corresponding peaks in
the H~I column density along the arm.

The brightest peak in UV emission ($12^{h}41^{m}31.7^{s}$
$+41^{\circ}08^{\prime}0.5^{\prime\prime}$) in NGC 4618 has a flux
comparable to that of a massive star cluster and is located on a wispy
H~I filament with a density of 7.3 M$_{\odot}$pc$^{-2}$. This ridge of
H~I can most easily be seen in Figure 8 separating holes 1 \& 2.  We
analyzed the {\it GALEX} observations and derived FUV and NUV fluxes
of 1477 $\mu$Jy and 970 $\mu$Jy, respectively, equivalent to
1.9$\times$10$^{-14}$ and 5.5$\times$10$^{-15}$ erg s$^{-1}$
cm$^{-2}$ {\AA}$^{-1}$ (-14.30 and -12.95 magnitudes, respectively).
These fluxes are a few times lower than the 9.8$\times$10$^{-14}$
erg s$^{-1}$ cm$^{-2}$ {\AA}$^{-1}$ measured in NGC 5471, a giant H~II
region in M101 (Garcia-Benito et al. 2011).  We conclude that this
source is likely a giant star cluster in NGC 4618 and may have
been responsible for the formation of H~I holes in the southern part
of the galaxy.

In the disk of NGC 4625, we see no 
radio continuum emission, but observe prominent UV emission arising from the
areas of highest H~I column density (e.g. the disk and brightest
spiral arm) as shown in Figure 11.  Gil de Paz et al (2005) studied
the UV emission in the disk of NGC 4625 in great detail and suggest
the possible existence of multiple spiral arms and large structures
associated with the galaxy. They also point out that the UV emission
extends roughly four times farther than the stellar disk as measured
by $D_{25}$ (Figure 3 of Gil de Paz et al. 2005).  Gil de Paz et al
(2005) reason that the UV emission at the large radii is likely due to
gravitational instability in the outer H~I disk. They further suggest the likelihood of an unknown additional factor that is affecting only the disk of NGC 4625 in order to account for the lack of extended UV emission in NGC 4618. If the extendended UV emission is indeed due to a gravitational instability in the disk, it would have to be caused by an current interaction with NGC 4625A, due to the lack of an interaction between NGC 4618 and NGC 4625, as we argue in $\S$5. It is difficult, however, to say to what degree the proposed dwarf galaxy could perterb the disk of the significantly more massive NGC 4625.
What we see in Figure 11 is that the H~I disk of NGC 4625 is quite
extended and that the UV emission falls along enhancements in the H~I
column density that morphologically resemble spiral arms.  Gil de Paz
et al.  (2005) suggest that there is faint UV emission associated with
what we now see as the outermost H~I arm.

\subsection{Star Formation Thresholds and H~I Column Densities}

Kennicutt (1989) argued that large scale star formation in disk
galaxies only occurs once the surface density of gas crosses a certain
threshold following the global definition of the critical density,
\begin{center}
{$$\Sigma_{c}=\alpha\frac{\kappa c}{3.36G}$$}
\end{center}
where $\kappa$ is the epicyclic frequency,
\begin{center}
{$$\kappa=1.41\frac{V}{R}(1+\frac{R}{V}\frac{dV}{dR})^{1/2}$$}
\end{center}
c is the velocity dispersion and $\alpha$ is a dimensionless constant
often taken to be less than 1 due to the existence of
fluid instabilities in a gas and stellar disk (Jog \& Solomon
1984\textit{a,b}).  While it is clear that the more accurate assessment 
of the critical density requires knowledge of the molecular content (e.g.
Schruba et al. 2011), we can compare the surface density of atomic
gas and the critical density with our existing data.  We adopt a value for
$\alpha$=0.67, which is the mean value for a sample of disk galaxies
as derived by Kennicutt (1989). The value for the velocity dispersion
of the gas, {\it c}, was taken to be the measured average value from the
respective regions on our velocity dispersion map (Figure 5).

Using the above mentioned values and those listed in Table 2, we
calculate the star formation threshold for the distinct areas in each
galaxy. For NGC 4618, the disk has a calculated critical density of 40
M$_{\odot}$pc$^{-2}$, whereas the ring has a much lower threshold at
6.5 M$_{\odot}$pc$^{-2}$. This large difference is the result of the
steep rise in the rotation curve in the central parts of the galaxy
while the rotation curve in the outer ring is flat.  The threshold in
the center of NGC 4625 is 39 M$_{\odot}$ pc$^{-2}$ and $\sim11$ M$_{\odot}$
pc$^{-2}$ in the spiral arm.

\begin{table}[ht]
\centering
\caption{Values for Critical Density Calculations}
\begin{tabular}{c c c c c c c}
\hline\hline
Galaxy & Region & Radius & Velocity & c & $\frac{dV}{dR}$ & $\Sigma_{c}$ \\
\hline
& & (kpc) & (km s$^{-1}$) & (km s$^{-1}$) & ($\frac{km/s}{pc}$) & (M$_{\odot}$ pc$^{-2}$) \\
[0.5ex]
\hline
NGC 4618 & Disk & 2.15 & 80 & 15.3 & .0477 & 39.97 \\
NGC 4618 & Ring & 5.82 & 100 & 7.0 & 0 & 6.47 \\
NGC 4625 & Disk & 1.42 & 55 & 10.0 & .0423 & 39.18 \\
NGC 4625 & Stellar Arm & 2.65 & 55 & 7.3 & 0 & 10.79 \\
[1ex]
\hline
\end{tabular}
\end{table}

With the motivation of understanding the global star formation
thresholds in both galaxies, we derive corresponding H~I column
densities for the distinct areas in both galaxies, using beam size and
average flux per beam as stated in $\S$2. The observed peak surface
density in the disk of NGC 4618 (Figure 10) is $\sim21$
M$_{\odot}$pc$^{-2}$, a factor of two below the nominal critical
density. The ring surrounding NGC 4618, which is not seen to have
on-going star formation, is measured at having an average density of
of 0.9 M$_{\odot}$pc$^{-2}$, well below the calculated critical
density.

The surface density of the atomic gas coincident with the optical disk
of NGC 4625 is $3.9$ M$_{\odot}$pc$^{-2}$, falling to $\sim$3.0 M$_{\odot}$pc$^{-2}$ 
for a prominent spiral arm of NGC 4625, as seen in Figure 11.  The full
radial plots of the H~I surface density for both galaxies are shown in
 Figure 12.  These profiles are consistent with the set of radial
profiles for another sample of dwarf and irregular galaxies discussed by
Wilcots \& Hunter (2002).

Gil de Paz et al. (2005) note the existence of the extended UV disk in
NGC 4625 and sketch the spiral UV morphology of the disk. The measured
H~I column density in the outer arms is well below the nominal
critical density, however the spiral arms correspond to regions in the disk
where the H~I peaks and where we observe UV emission and therefore, on-going star formation. In NGC
4625, on-going star formation can be observed to a distance
$\frac{3}{4}$ the radius of the H~I disk (BW04), which is also a
distance 3 times greater than the radius of the optical disk.  The
fact is that in both galaxies star formation proceeds even when the
surface gas density of the atomic gas is below the nominal critical
density.  However, we do not include molecular gas and Schruba et
al. (2011), among others, have clearly demonstrated that the surface
density of star formation is not well correlated with the surface
density of atomic gas in disk galaxies.  Our results also reinforce
the idea that in many galaxies, the onset of star formation is a
result of local events rather than a global instability in the disk.

\subsection{Velocity Dispersions of the Disks}

The mechanical energy deposited by massive stars into their environment is
typically more than sufficient to account for the formation of H~I holes, 
especially when integrated over several generations of stars (e.g. Warren
et al. 2011).  Thurow \& Wilcots (2005) and Wilcots \& Thurow (2001) also
showed that there is more than enough energy to account for flows of ionized
gas in irregular galaxies.  The mechanical energy from feedback could also go
into heating the cool ISM, resulting in higher H~I velocity dispersions 
around regions of recent star formation.  Such correlations are common in 
star-forming galaxies (e.g. IC 10 [Wilcots \& Miller 1998], NGC 1569 [Muehle et al.
2005, Walter et al. 2008]).

The H~I velocity dispersion in NGC 4625 is nearly uniform across the entire
disk at a value of $\sim10$ km s$^{-1}$ with little variation between regions with
and without UV emission.  Combined with the lack of noticeable H~I holes,
this suggests that the current and recent star formation in NGC 4625 is having little
or no impact on the neutral ISM in the galaxy.  

The situation in NGC 4618 is quite different.  Not only do we see H~I
holes, we also see a significant increase in the velocity dispersion
of the gas in central regions coincident with the current star
formation. We see a peak velocity dispersion of ~30 km s$^{-1}$
located between the peak in the UV emissions (which we discussed in
$\S$4.2) and a star-forming region in the stellar arm.  The variation 
in the velocity dispersion throughout the disk is
largely coincident with peaks in the H~I column density and ongoing
star formation. There are also areas of higher velocity dispersion in
the SE and NW corners of the disk where it appears that the disk intersects
with the H~I ring discussed below. In these regions, the H~I profiles are
noticeably wider when compared to other sections of the ring.

\section{Are NGC 4618 and NGC 4625 Interacting?}

The notion that NGC 4618 and NGC 4625 are interacting is conveniently
consistent with the idea that the asymmetric nature of Magellanic spirals 
is the result of an interaction with a close neighbor (e.g. Odewahn 1994).
The apparent proximity of the galaxies to one another and the apparent
overlapping H~I distributions are highly suggestive of an interaction. 
BW04 stated that any interaction between NGC 4618 and NGC 4625 would
only be affecting the outermost H~I of the galaxies -- this led them to
believe that the observed asymmetry in the stellar disk of NGC 4618 is not a
consequence of the current interaction unless the response of the
disks is abnormally rapid.  The lack of perturbations seen in the
velocity field of NGC 4625 is further indication against any on-going
interaction between the two galaxies. In lue of an on-going interaction, the extended UV emission observed in NGC 4625 cannot be due to gravitational instabilities, as suggested by Gil de Paz et al. (2005). The H~I profile asymmetries
measured in NGC 4618/4625 are on the order of the asymmetries measured
in noninteracting Magellanic spirals (Wilcots \& Prescott 2004). 

Investigating the H~I profiles associated with the region where the galaxies'
H~I disks seem to overlap reveals two distinct kinematic components, 
separated by as much as 50 km s$^{-1}$ (Figure 13). If the galaxies were
interacting we would expect to see a smooth gradient in the velocity
of the gas ``bridge." However, we do not see this, thus the galaxies
are likely separated and are not interacting. The lack of broadened
profiles or double peaks is further evidence against mixing of the
outer disks of the galaxies.

The true separation between the two galaxies is difficult to measure
due to the uncertainty in the orientations of the two galaxies
relative to each other and the lack of distance indicators that are
independent of the local gravity field. However, applying the
Tully-Fisher relation, Tully (1988) derived distances of 7.3 and 8.2
Mpc for NGC 4618 and NGC 4625, respectively.  The claim that there is
a 0.9 Mpc separation between the redshift-independent distances is consistent
with our conclusion that the apparently overlapping H~I disks are 
actually kinematically distinct and the NGC 4618 and NGC 4625 are not
currently interacting.

 \section{The Nature of the Ring around NGC 4618}

The presence of the H~I ring around NGC 4618 is a bit of a mystery.
Galaxies with extended H~I gas have been known to have H~I ``streamers" stemming
from the disk of the galaxy and wrap around the galaxy either
completely or in part (Hunter et al. 1998; van der Hulst 1979;
Thomasson \& Donner 1993; Noguchi 1988). These streamers are often
the consequence of an external perturbation, particularly an interaction
with a dwarf companion or massive gas cloud (e.g. Athanassoula \& Bosma 1985).  
It is possible that NGC
4618 does indeed have streamers that are feeding the ring, but at
our inclination angle we are observing them face-on. Perhaps NGC 4618
was once like NGC 4449 (Hunter et al 1998), but has long been able to
evolve a stable, complete ring, assuming the interaction causing the
formation of the ring was not recent.

An investigation into the overall kinematics and timescales of the H~I
ring surrounding the disk of NGC 4618 gives us some idea of a minimum
age of the ring and when NGC 4618 had a previous interaction. For
simplicity's sake, we assume that the H~I ring around NGC 4618 is
circular; therefore, we measure from our rotation curves a v$_{circ}$
of 95 km s$^{-1}$, which is the average rotational velocity of the
ring. Calculating the orbital speed out to 8.5 kpc, we calculate an
orbital period of 456 Myr.  The average velocity dispersion across the
ring is 6.8 km s$^{-1}$, which translates into an expected width of
$\sim$5 kpc after 500 Myr of diffusion, assuming a constant velocity
dispersion.  The maximum velocity dispersion is 36 km s$^{-1}$, which
would lead to a width of 18.4 kpc.  The observed width of the ring is
$\sim$3.7 kpc, suggesting that it is either much younger than an
orbital period or is at least 450 Myr old and is stable enough to hold
itself together.

If the ring is the result of an interaction, it is not at all clear
with which galaxy NGC 4618 could have interacted. Using the difference
between their systemic velocities and a 900 kpc separation (Tully
1988) between the two galaxies, the minimum crossing time of NGC 4618
and NGC 4625 is 1.6$\times$10$^{10}$ years, longer than the age of the
Universe.  In order to assess the possibility that either NGC 4618 or
4625 has had a different recent encounter, we used the Sloan Digital
Sky Survey (SDSS) DR7 (Abazajian et al. 2009) to search for nearby
objects. The closest neighbor is NGC 4490, a spiral galaxy also known
as the Cocoon galaxy, with a v$_{sys}$=565 km s$^{-1}$ and an angular
separation of 2.74$^{\circ}$ from NGC 4618.  At a distance of 7.8 Mpc
(Tully 1988) this corresponds to a projected separation of 375 kpc on
the sky.  The difference between the systemic velocities of NGC 4618
and 4490 is only 21 km s$^{-1}$ which would imply a last encounter
$3.8\times10^{10}$ years ago.  However, we can say nothing about
their tangential motions relative to each other. In order for the two
galaxies to have crossed 450 Myr ago, NGC 4618 and NGC 4490 would have
to be moving at $\sim$820 km s$^{-1}$ relative to one another. While
this would be an unusually high relative velocity between two galaxies
not in a cluster, it is more reasonable to suggest that the ring has
survived for 2-3 orbit times and an encounter occurred 1-1.5 Gyr ago.

The ring does not necessarily have to be a remnant of an earlier
interaction with another galaxy of similar mass.  As in the case of
NGC 4618, the ring around NGC 4449 (Hunter et al. 1998) completely
encircles the central part of the galaxy at a position angle offset
from that of the inner regions.  Hunter et al. (1998) suggest that the
morphology of the ring around NGC 4449 is the result of an encounter
with a nearby dwarf companion.  However, aside from the hint of a
dwarf companion to NGC 4625 mentioned in Gil de Paz et al. (2005) we see 
no such companion near NGC 4618.  

At this point too little is known about this putative companion, NGC
4625A to say whether or not it is or could have interacted with either
NGC 4625 or NGC 4618. The H~I disk of NGC 4625 is remarkably quiescent
with a nearly uniform velocity dispersion and if the proposed
companion galaxy NGC 4625A exists, it is devoid of neutral gas within
our velocity range. Our one channel 5$\sigma$ detection for NGC
4625A is 1.26$\times$10$^{4}$M$_{\odot}$. This strongly suggests a
lack of neutral gas or, if there is gas present, it is not in our
velocity range and therefore not a likely companion of NGC 4625, nor
NGC 4618.  We cannot rule out the possibilty that NGC 4625A is a 
dwarf elliptical or spheroidal companion to NGC 4625.

We are left with the same issue that has long bedeviled our efforts to
understand the origin of the asymmetric morphology of Magellanic-type
galaxies.  Despite the numerous models that show how one can account
for the off-center bars and one-armed spiral, NGC 4618 (and most other
Magellanic galaxies) show no sign of a companion.  Noodermeer, Sparke,
\& Levine (2001) suggested that disk galaxies, once knocked off-center
remains so as long as the disk is much less massive than that the halo.
Those authors do not investigate what could have knocked the disk
off-center in the first place.  Besla et al. (2012) recently proposed
that the Magellanic Stream, and the structure of Magellanic galaxies
in general, could be explained with the merger of two dwarf galaxies.
The implication for our study would be that NGC 4618 is the product of
past merger of two galaxies.  There has also been a recent suggestion that
the characteristic asymmetric morphology of Magellanic spirals could
be the result of a collision with a dark satellite (Bekki 2009).  The
obvious attraction of this suggestion is that it gets around the
problem that most Magellanic spirals do not have companions.

\section{Conclusions}

We presented a detailed analysis of high resolution H~I observations of
NGC 4618 and NGC 4625.  In combination with tracers of recent star formation
these high resolution H~I data allow us to carry out an investigation 
of the relationship between massive star formation and the distribution and
kinematics of the atomic interstellar medium in both galaxies.  
While the surface density of the atomic gas remains well below the nominal
critical density needed for global star formation (e.g. Kennicutt 1989), we
do find a spatial correlation between tracers of current star formation
and relative peaks in the H~I column density.  In NGC 4625 star formation
traces faint spiral structure that is also seen in the H~I column
density.  In NGC 4618 star formation is found in much of the disk.

We also noted a number of H~I holes in NGC 4618, but also show that
the holes are not expanding, and establish anti-correlation between
the location of the H~I holes and peaks in UV emission for the galaxy.
As with a number of other studies we conclude that the major holes are
most likely the result of supernovae.  However, we reach this
conclusion without an understanding of the molecular content of the
holes, but believe that the holes are more likely to contain hot
(10$^{6}$ K) gas rather than cold gas.  The 1.4 GHz radio continuum
emission stemming from the center of NGC 4618 suggests that the
emission stems from multiple supernovae, strengthening the argument
that the holes were created through a series of powerful
explosions. We investigate the environments in which we observe active
star formation in both galaxies and measure H~I column densities in
the area of the brightest star formation corresponding to the areas of
brightest UV emission.

The H~I line profiles associated with the gas believed to be the
bridge connecting the two galaxies actually has two well-defined and
separate kinematic components. This observation makes a strong case
against an on-going interaction between the NGC 4618 and NGC 4625
galaxies, in conjunction with the lack of perturbations to either velocity
field or the H~I velocity dispersion in NGC 4625.
 
The persistence of the H~I ring around NGC 4618 allows for us to apply
a rough interaction timescale of 450 Myr, assuming that the ring has
successfully completed one complete rotation around the center of the
galaxy. Investigating the area that surrounds the NGC 4618/4625
system, we find that NGC 4490 is projected to be relatively close to
the pair, yet serves as an unlikely candidate for having previously
interacted with NGC 4618. Thus, the origins of the likely stable H~I
ring around NGC 4618 are still not well understood. 

\section{Acknowledgements}
We thank Cody Gerhartz for helpful conversations and contributions to
this paper and acknowledge the support of NSF Grant AST-0708002.  We also
thank the anonymous referee for helpful comments to improve this paper.

\clearpage

\begin{figure}[h*]

\begin{minipage}[h*]{1.00\linewidth}
\begin{tabular}{l l}
\includegraphics[width=0.5\linewidth]{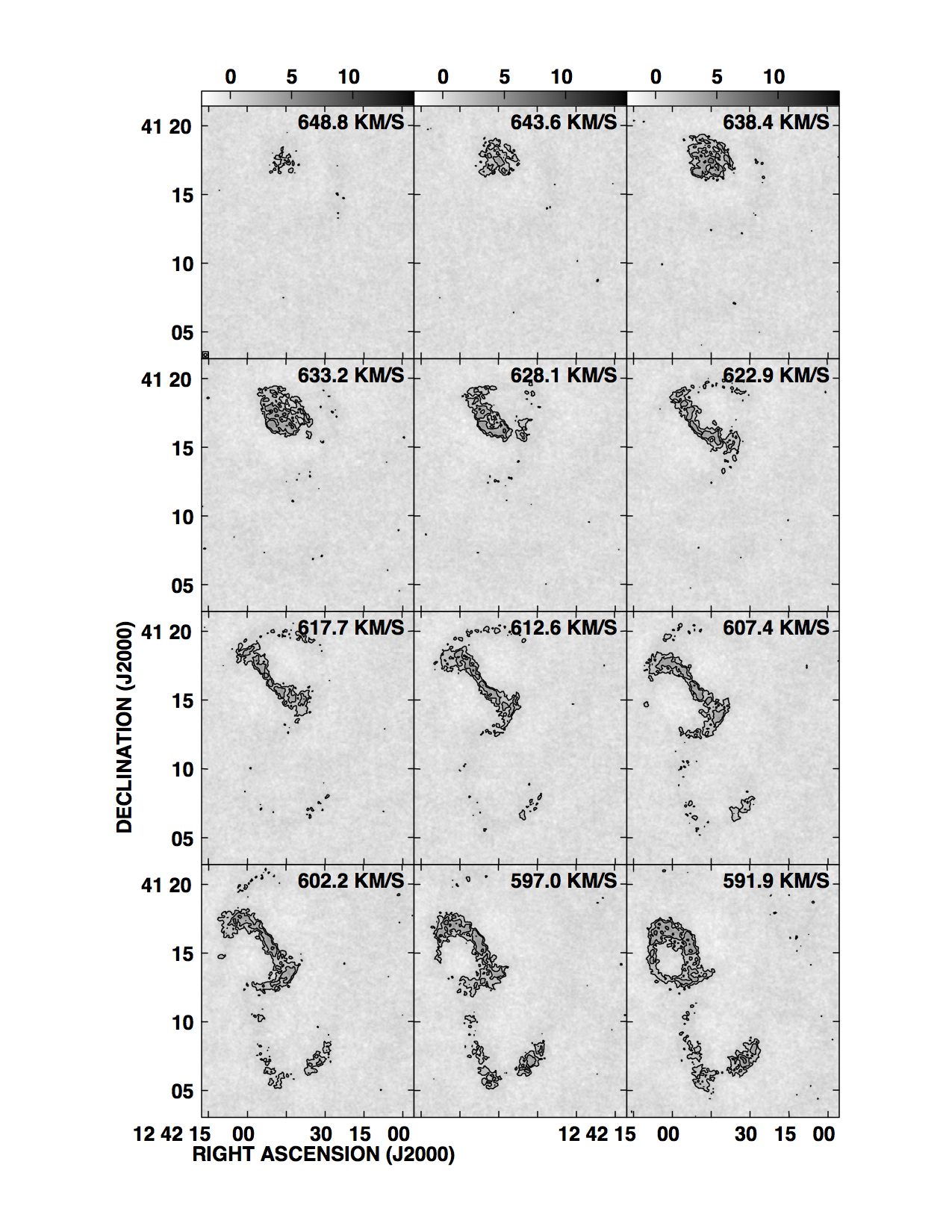} &
\includegraphics[width=0.5\linewidth]{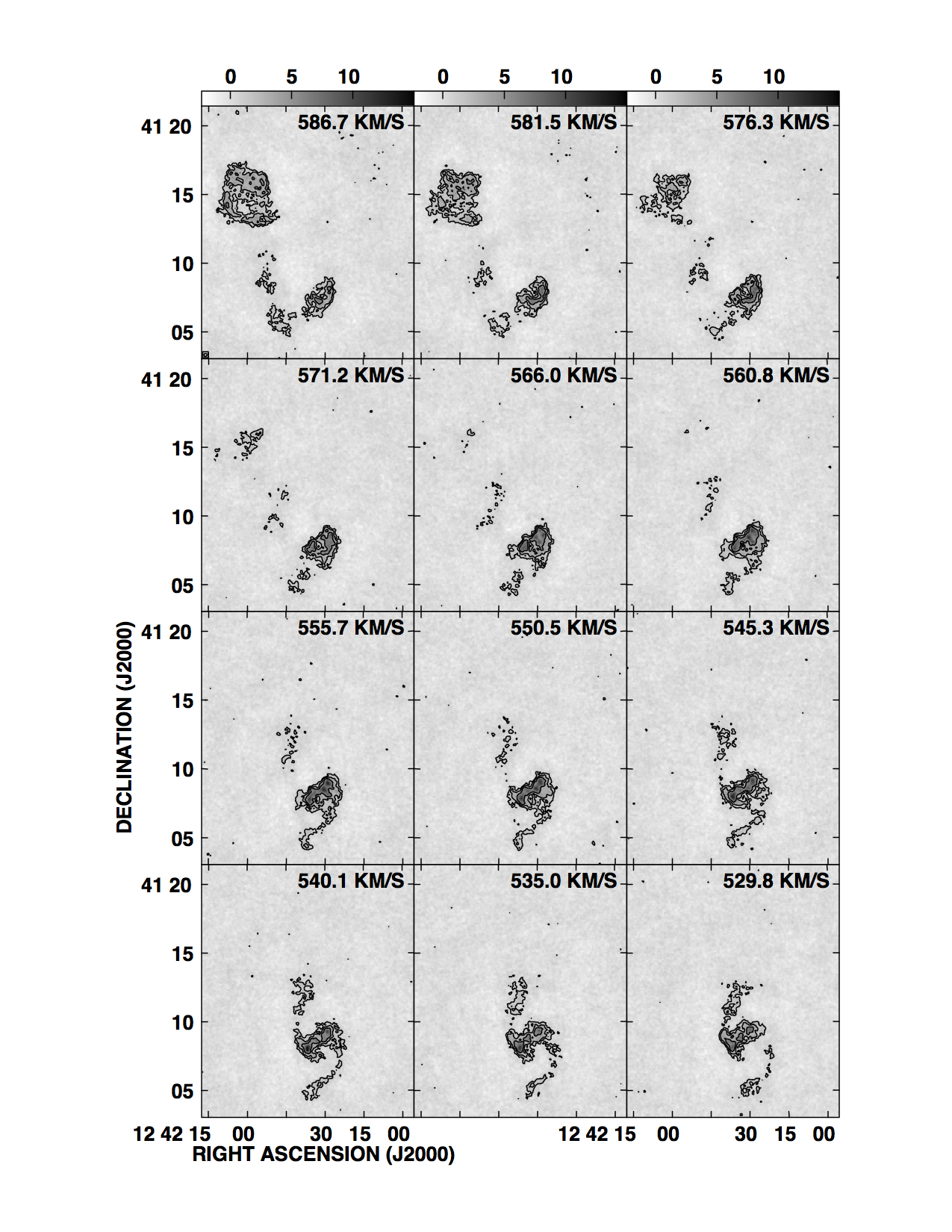}\\
\includegraphics[width=0.5\linewidth]{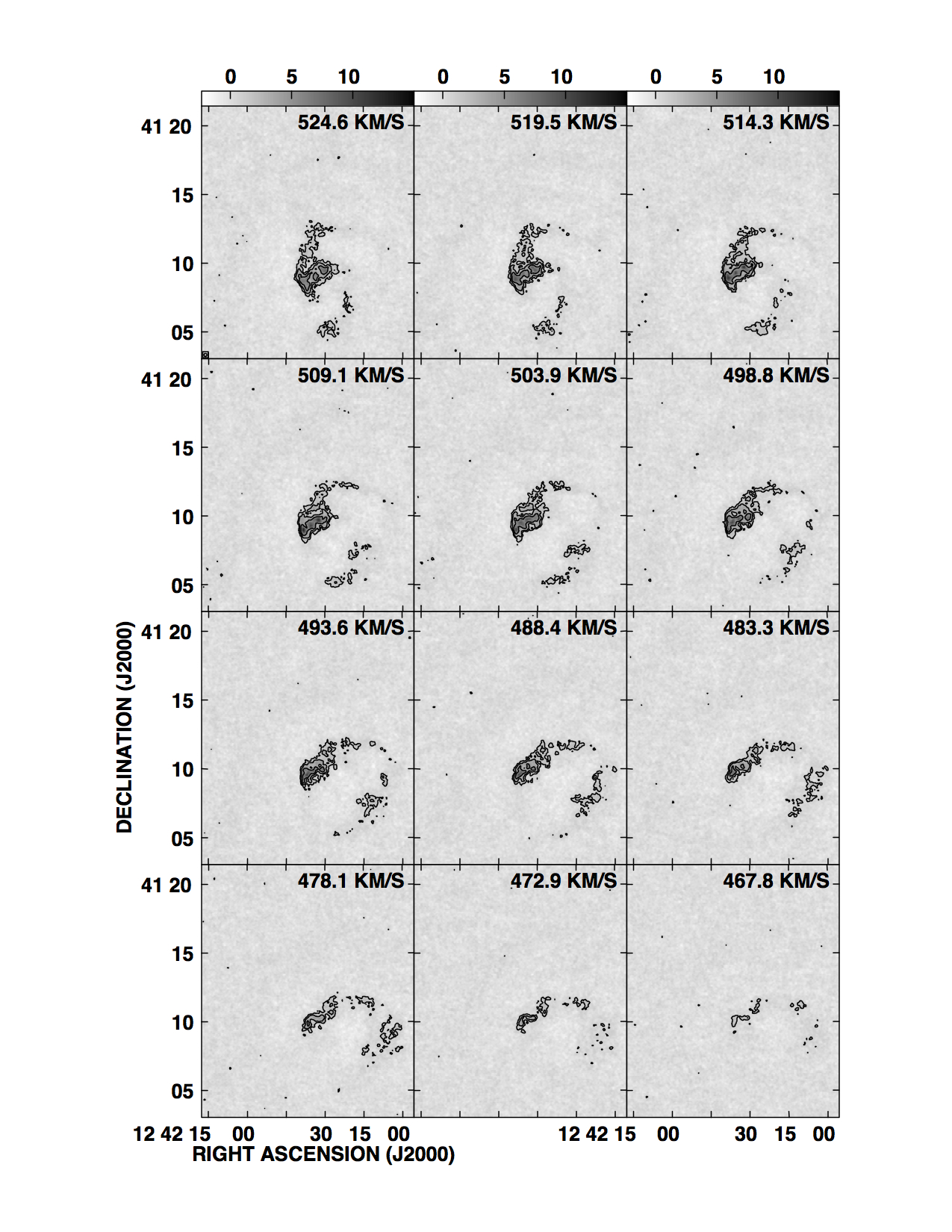} \\
\end{tabular}
\caption{\small Emission in each channel for the naturally weighted
data cube. Every channel with signal is shown. The contours are 3, 6 and 12
times the 2$\sigma$ noise level. NGC 4625 first appears in the upper left panel of the top left figure at a velocity of 648.8 km s$^{-1}$, whereas the initial emission from NGC 4618 appears at a velocity of 617.7 km s$^{-1}$. The beam is plotted in the bottom left corner of the top left frame in each image.} 
\end{minipage}

\end{figure}

\clearpage

\begin{figure}
\includegraphics*[angle=270, width=1\textwidth]{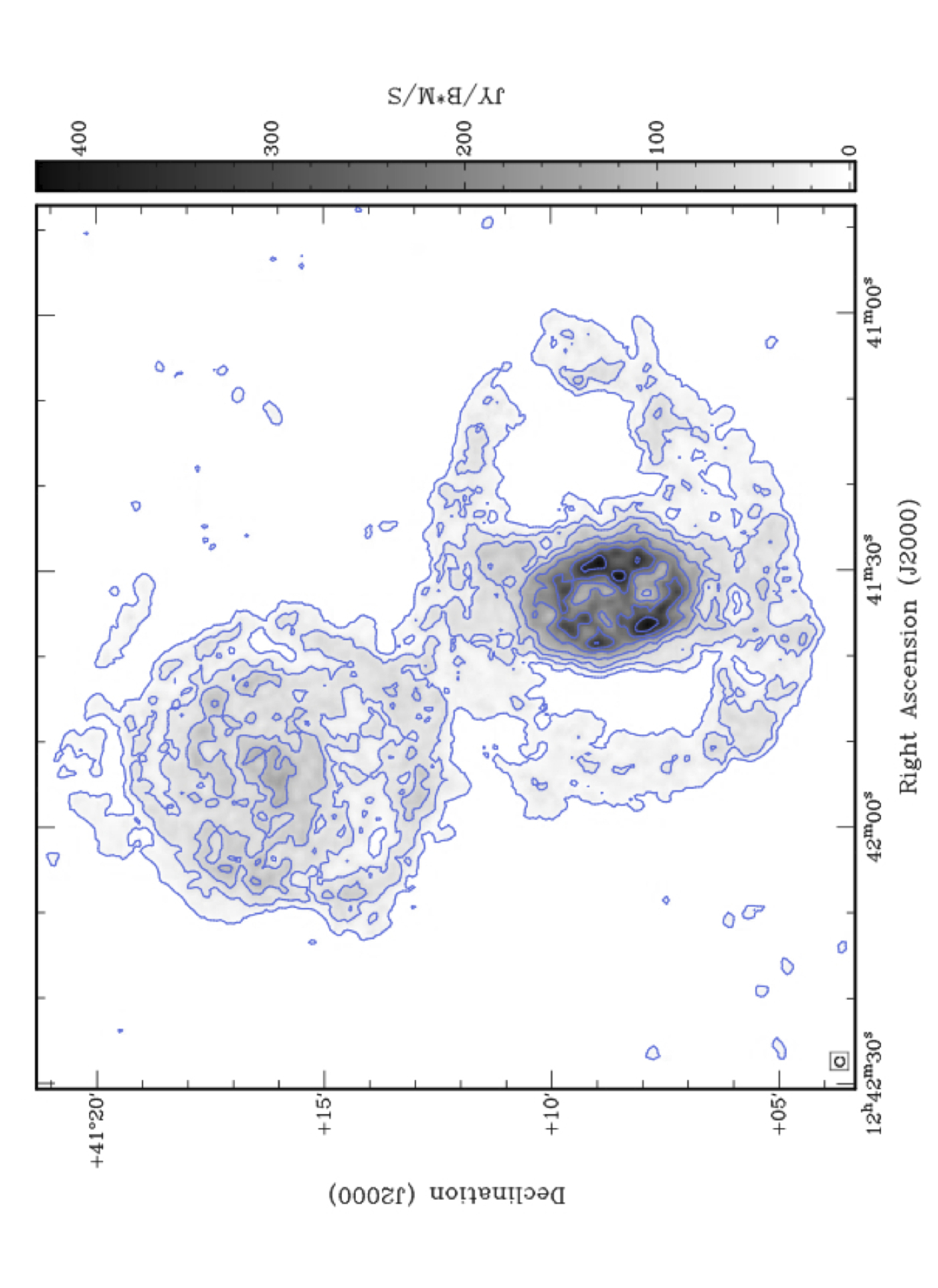}
\caption{\footnotesize Naturally weighted integrated-intensity map of NGC 4618
and NGC 4625. With a peak density of 21.4 M$_{\odot}$pc$^{-2}$, the contour lines represent 0.43, 1.7, 3.2, 5.4, 8.5, 12.8 and 17.1 M$_{\odot}$pc$^{-2}$. The beam is plotted in the lower left corner for reference.  The contrast between
the peak column densities in each galaxy stands out.  We note the existence a several H~I holes in NGC 4618
while none exist in NGC 4625.  It is also clear that NGC 4618 is encircled by a substantial ring of H~I gas which
is discussed in more detail in section 6.}
\end{figure}

\begin{figure}
\includegraphics*[width=1\textwidth]{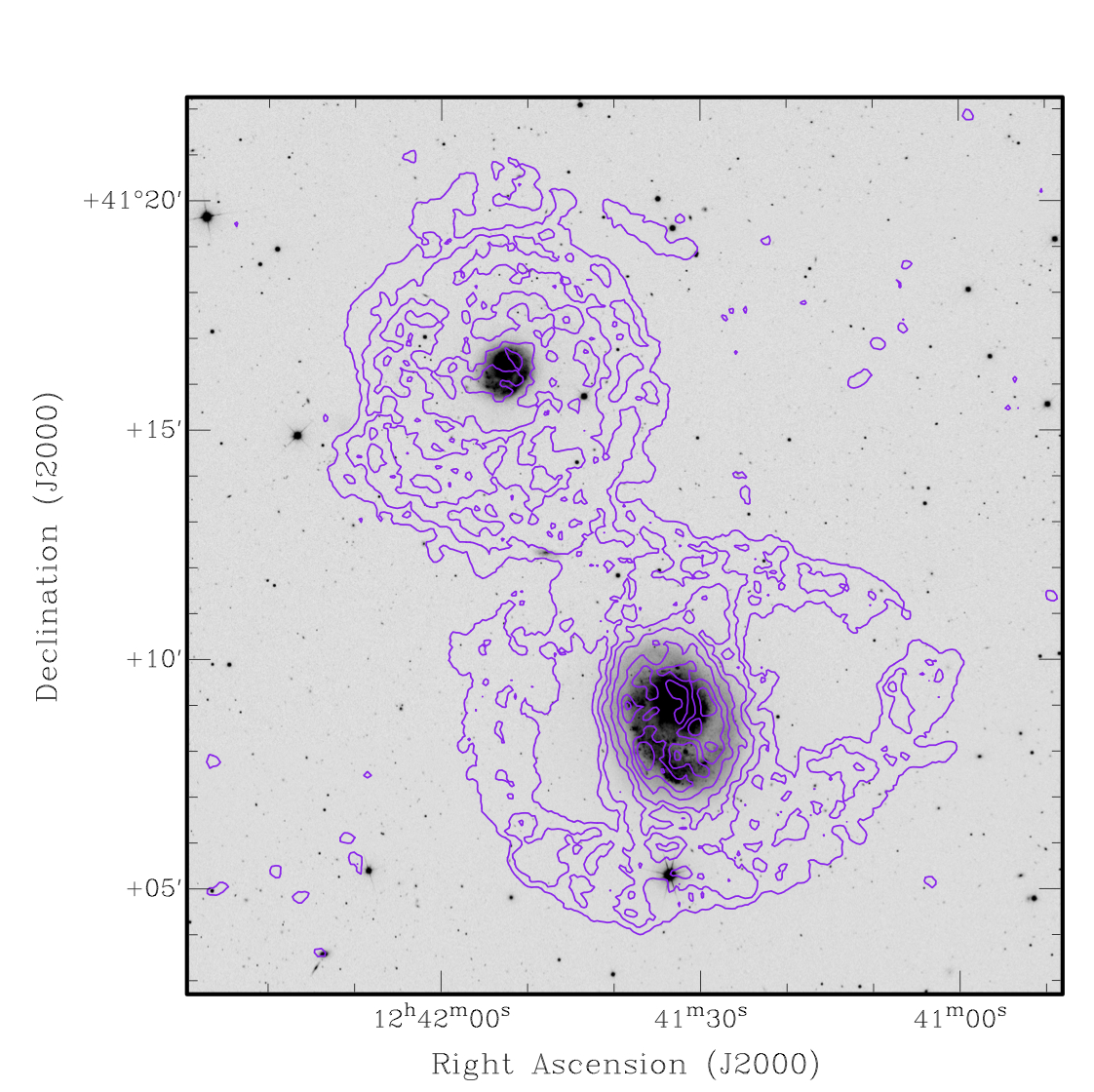}
\caption{\footnotesize Integrated H~I map overlaid on the SDSS r-band image of the NGC 4618 \& NGC 4625 field. The H~I contours represent the same intensity levels as Figure 2. The H~I beam is the same as that shown in Figure 2. We note in NGC 4625 the large H~I to optical disk ratio of 9.8 (Bush \& Wilcots 2004)}
\end{figure}

\begin{figure}
\includegraphics*[width=1\textwidth]{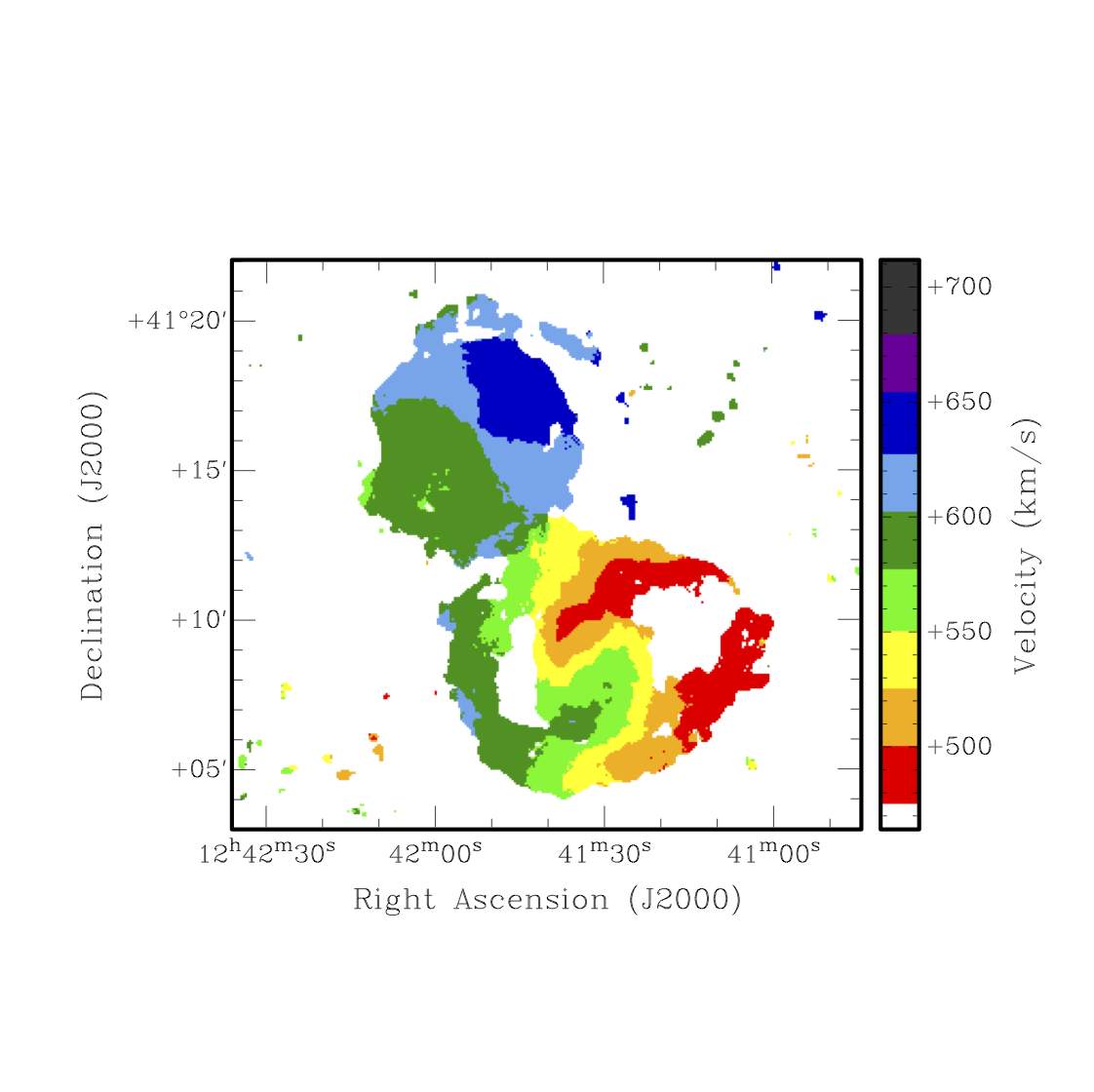}
\caption{\footnotesize Velocity map of naturally weighted data cube of
NGC 4618 and NGC 4625. The contours denote a velocity change of 10 km
s$^{-1}$. NGC 4625 typifies a differentially rotating disk, whereas NGC 4618 displays an ``S'' 
shaped isovelocity curve in the area associated with the stellar bar.}
\end{figure}

\clearpage

\begin{figure}
\includegraphics*[width=1\textwidth]{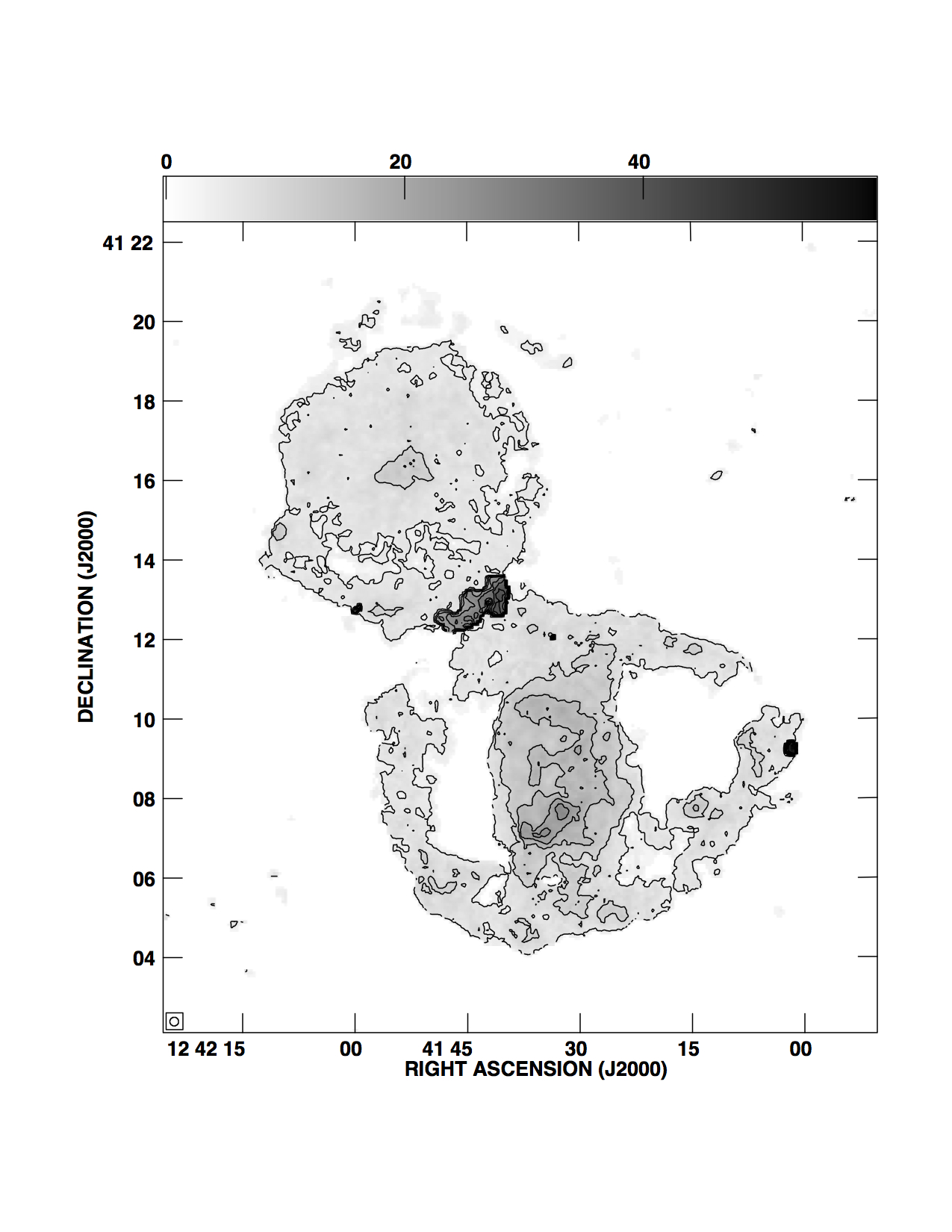}
\caption{\small Velocity dispersion of naturally weighted
data cube of NGC 4618 and NGC 4625. The area of a potential overlap
correlates to the area of the largest velocity dispersion. In NGC
4625, we see a remarkably constant velocity across the disk with the
exception of the central region. The
peaks in velocity dispersion in NGC 4618 are within the stellar extent
of the galaxy.}
\end{figure}

\clearpage

\begin{figure}
\begin{tabular}{c c}
\includegraphics*[width=.5\textwidth]{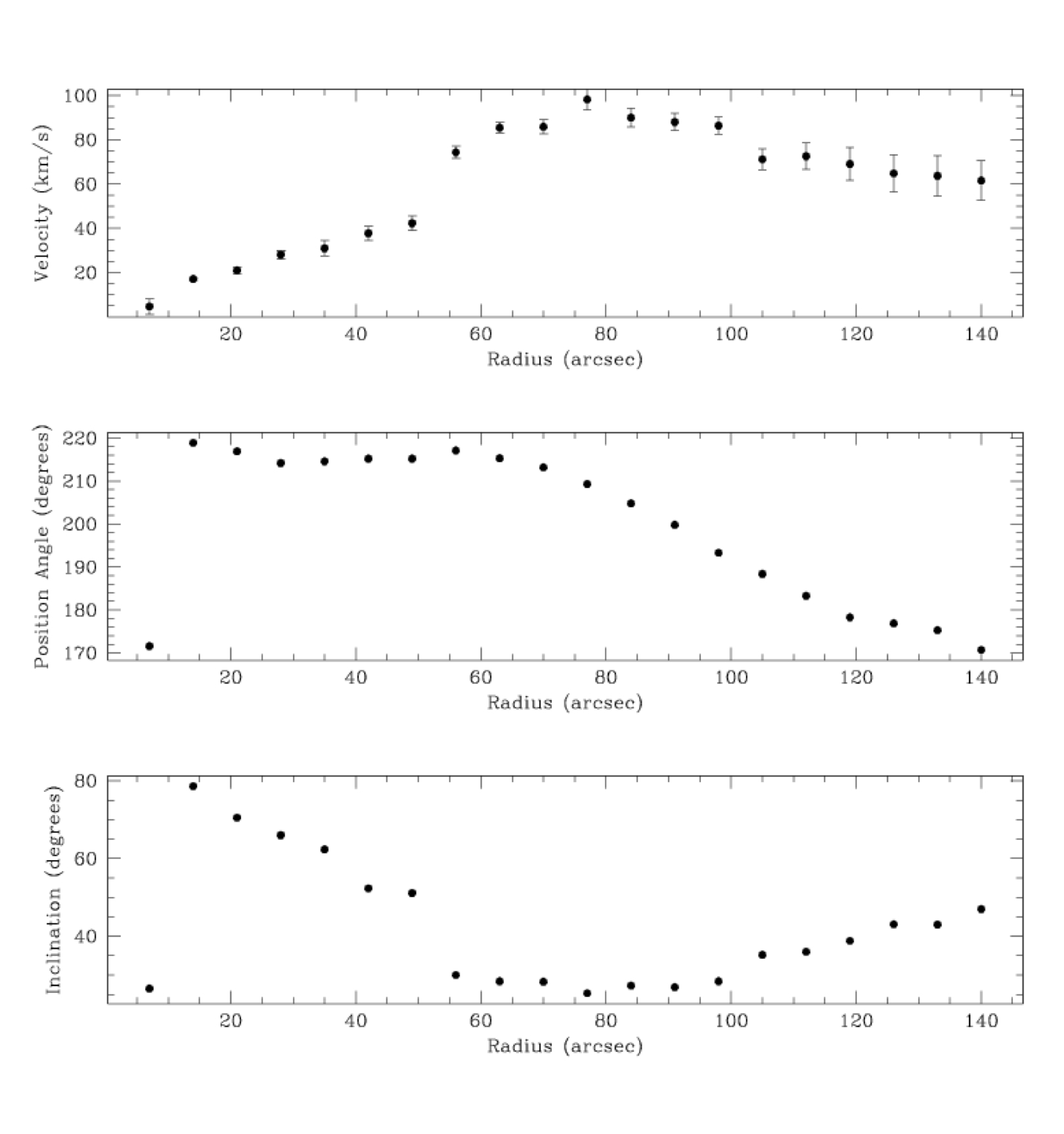} &
\includegraphics*[width=.5\textwidth]{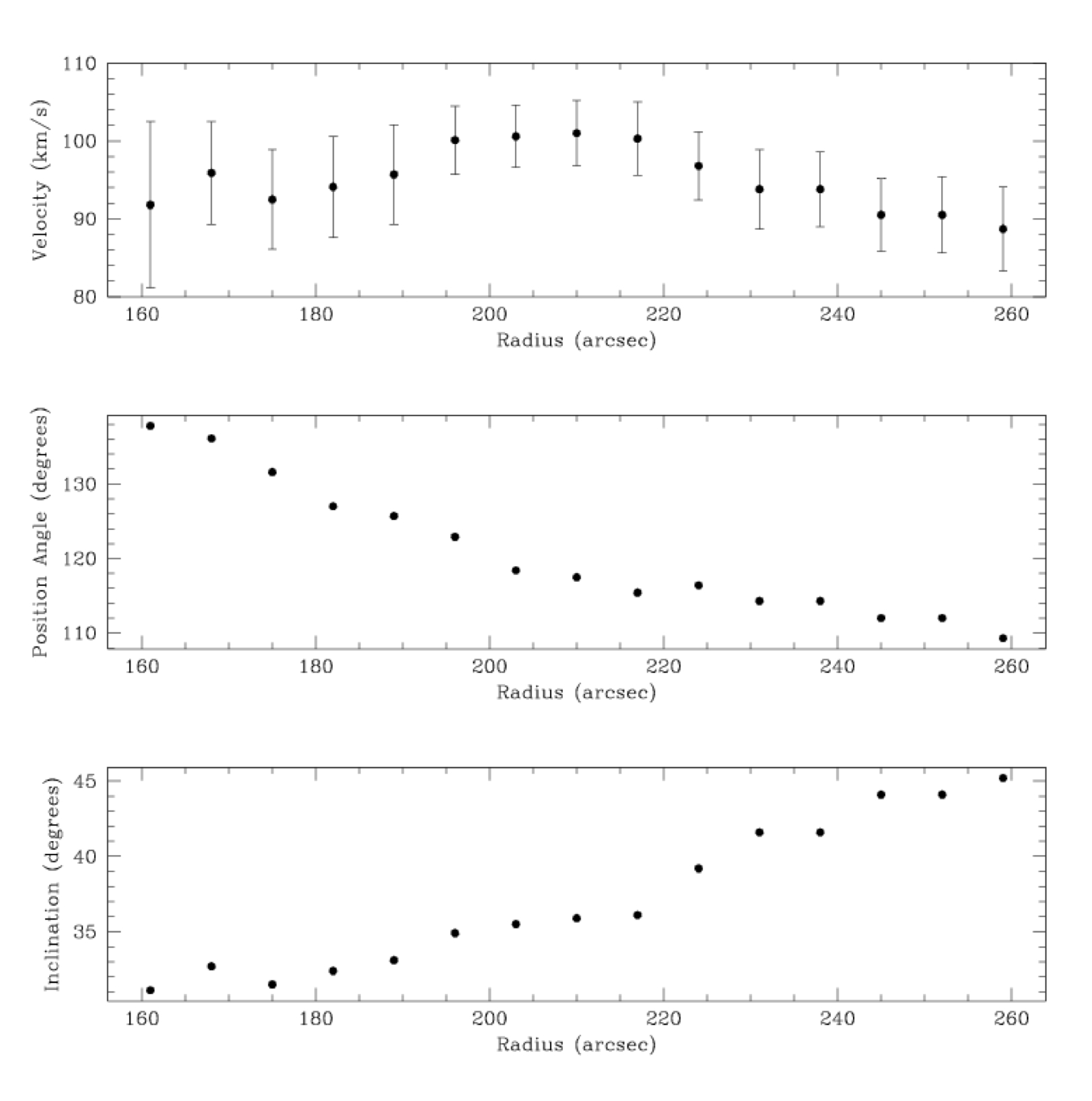} 
\end{tabular}
\caption{\footnotesize Rotation curve, position angle and inclination representing the optical disk \textit{(left)} and the H~I ring \textit{(right)} for NGC 4618. We integrate using 7$^{\prime\prime}$ rings. We note the shallow nature of the inner rotation curve \textit{(top panel)}. We can see a large jump from the disk to the ring of the galaxy in both position angle and inclination.}
\end{figure}

\begin{figure*}
\begin{tabular}{c c}
\includegraphics[width=.5\textwidth]{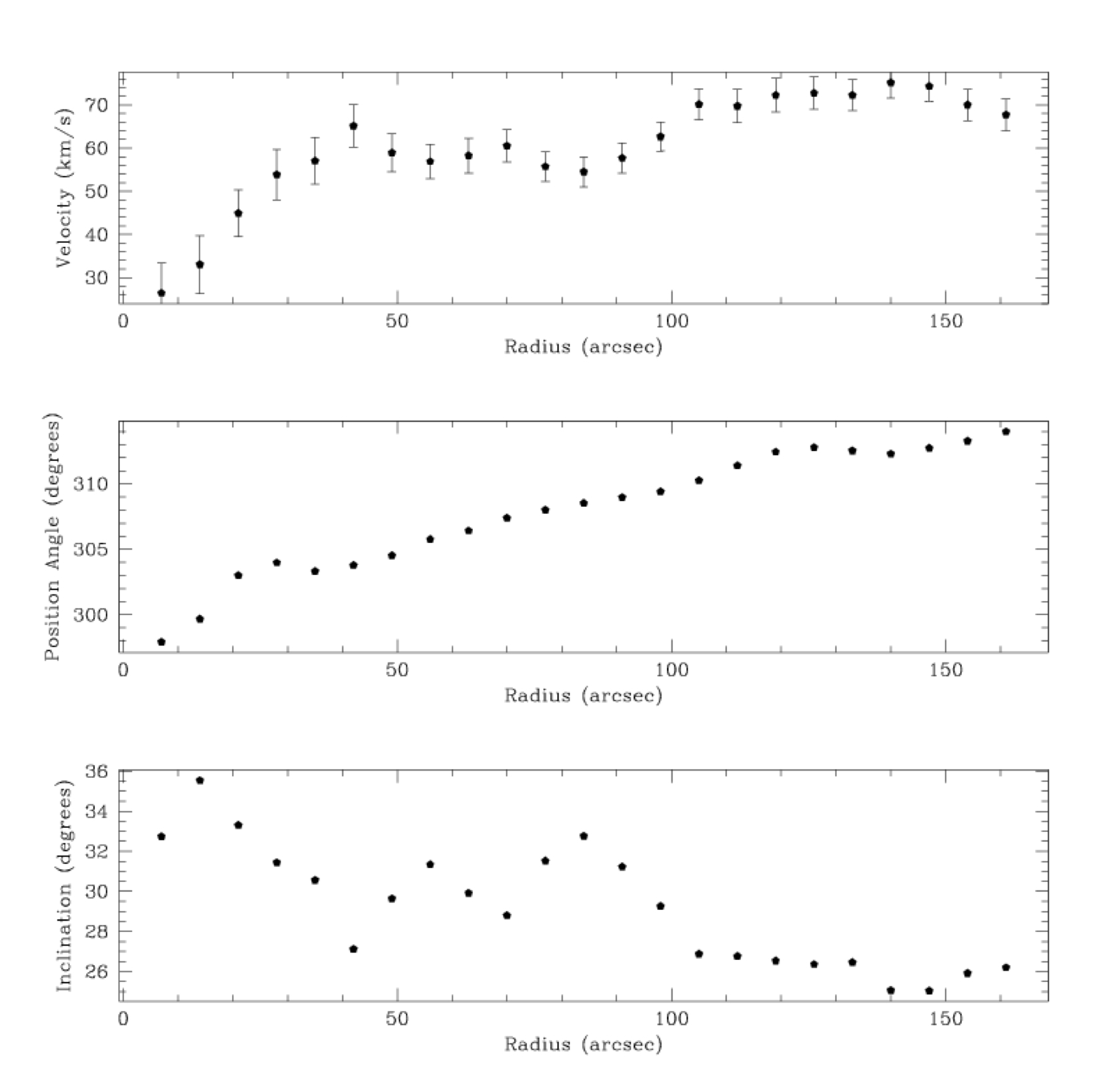} &
\includegraphics[width=.5\textwidth]{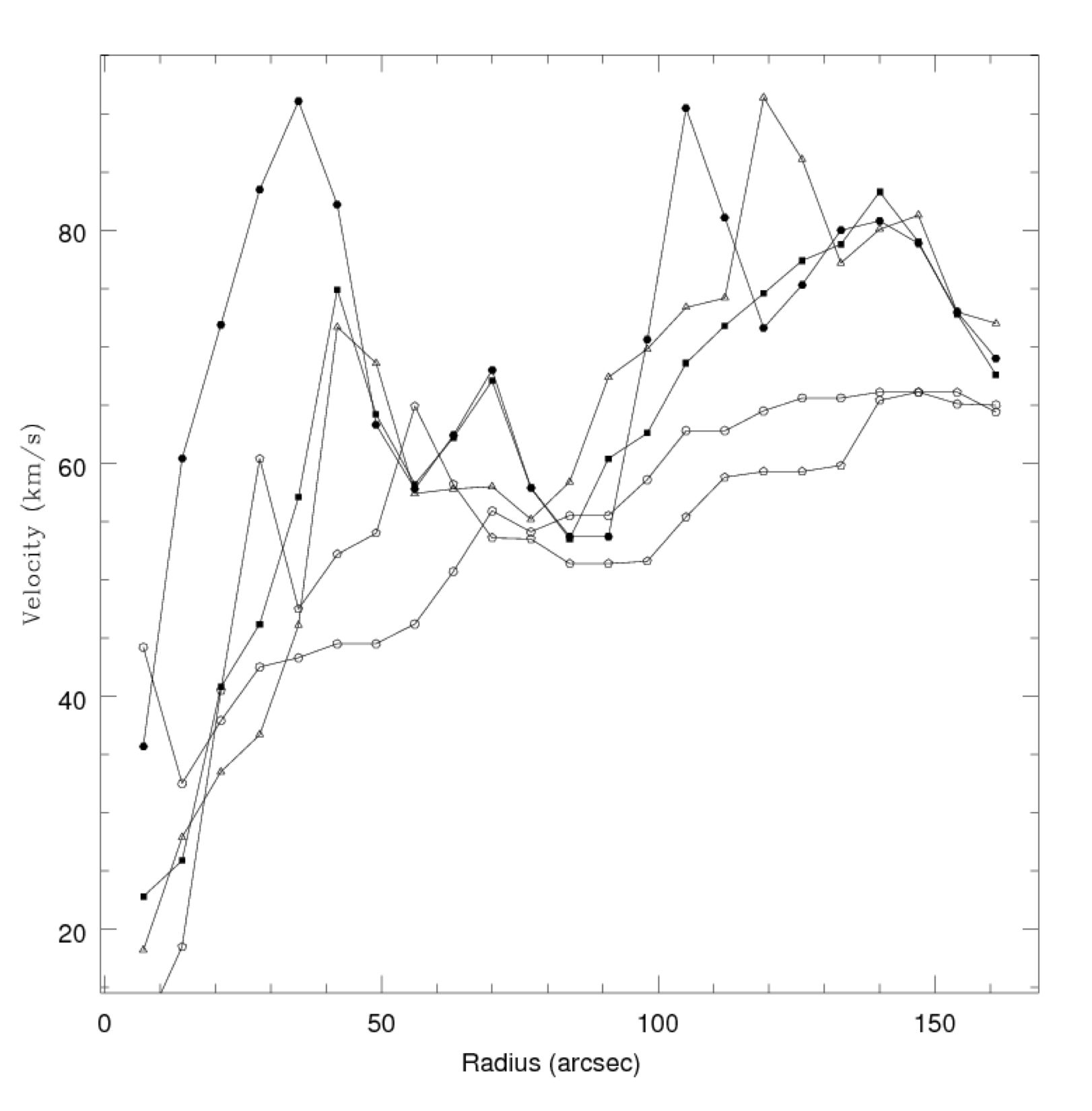}
\end{tabular}
\caption{\footnotesize (\emph{Top}) Rotation curve, position angle
and inclination corresponding to the averaged results from 6 solutions
of \emph{`gal'} for NGC 4625. Just as with NGC 4618, we integrate using 7$^{\prime\prime}$ rings. (\emph{Bottom}) The six separate solutions to the average
rotation curve corresponding to each running of \emph{`gal'}.}
\end{figure*}

\clearpage

\begin{figure}
\includegraphics[width=.9\textwidth]{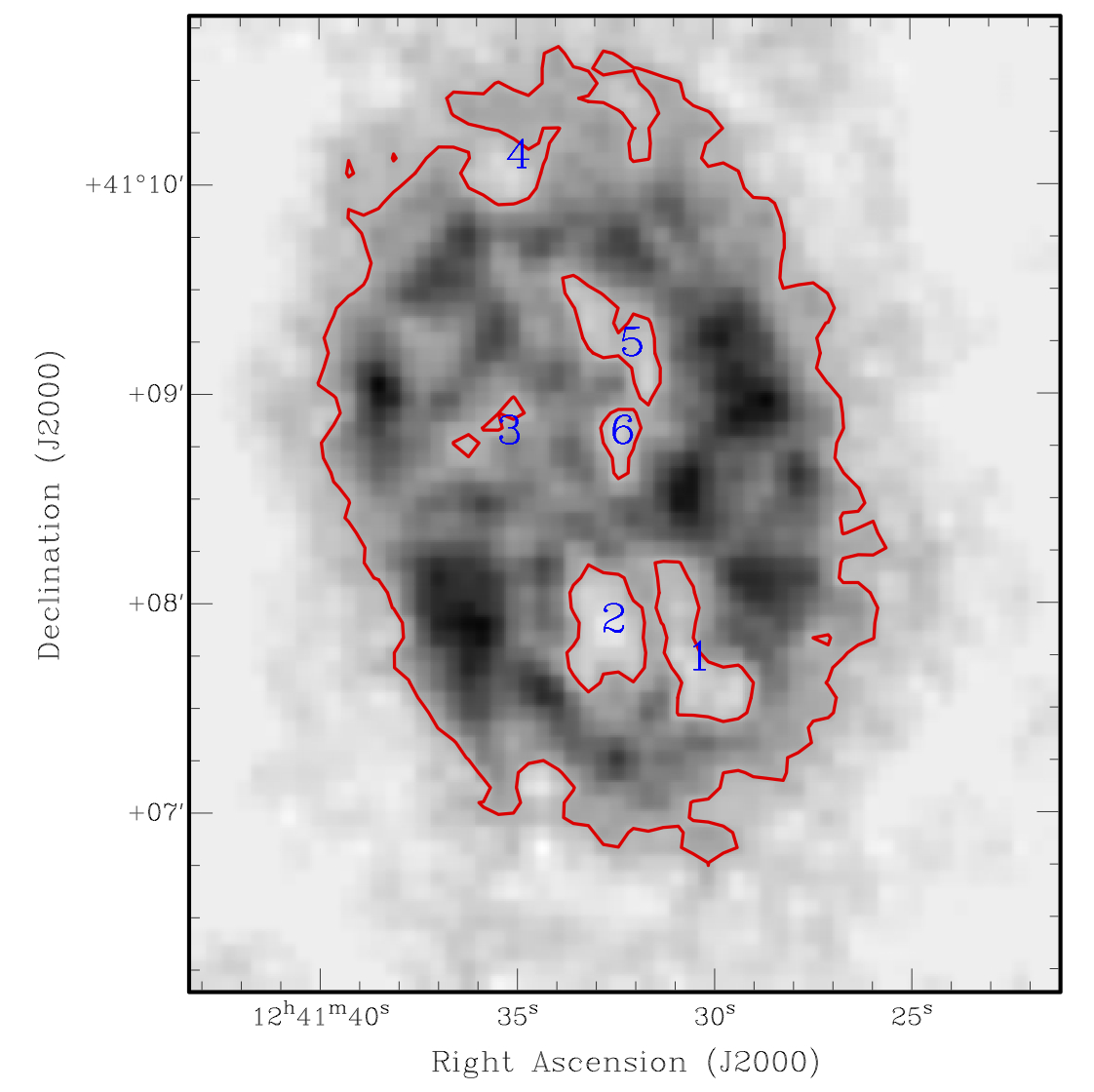}
\caption{Robust 0 moment 1 map with the six H~I holes having column densities less than 6.42 M${\odot}$pc$^{-2}$.
labeled.}
\end{figure}

\begin{figure}
\centering
\includegraphics[width=.6\textwidth, angle=270]{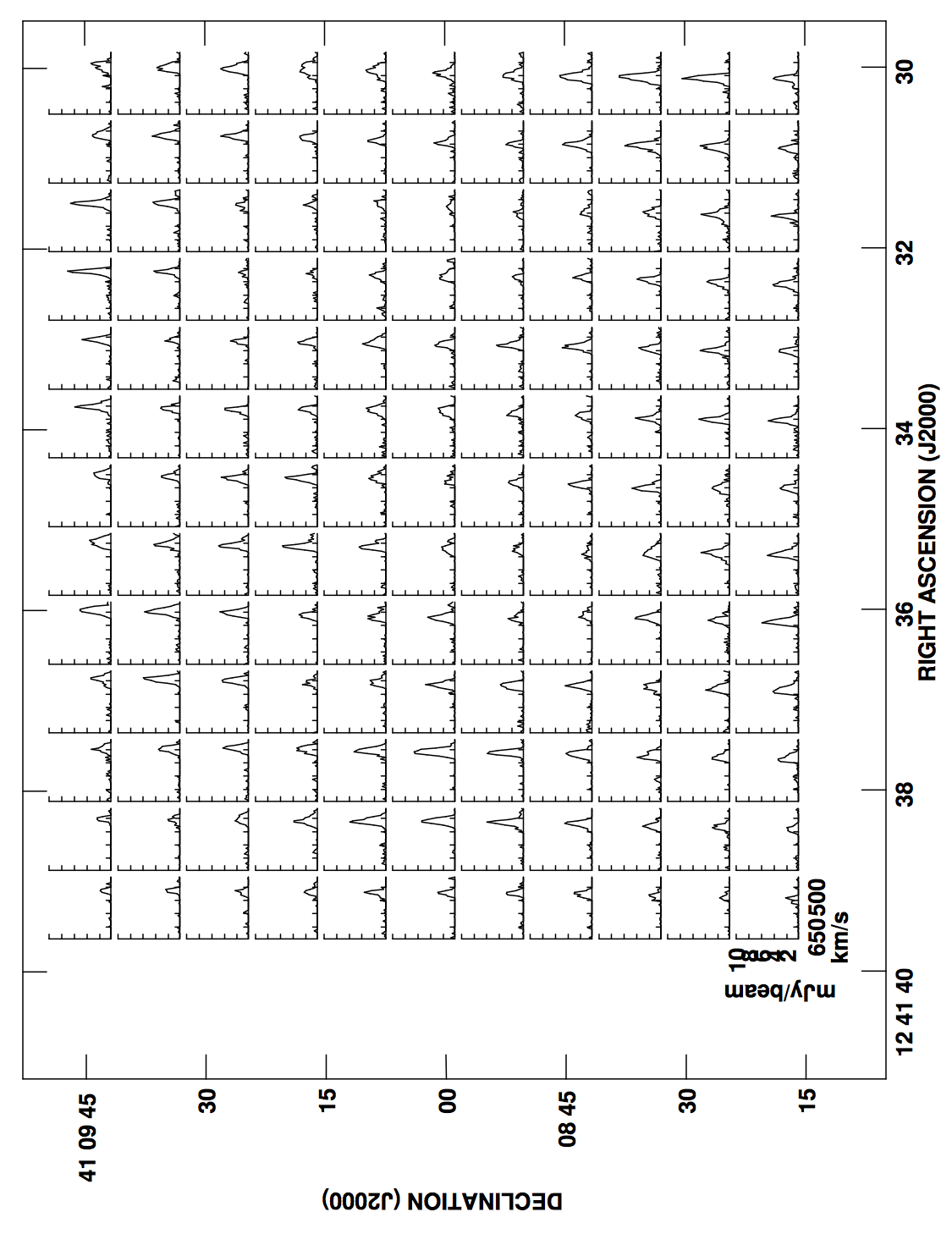} \\
\includegraphics[width=.55\textwidth, angle=270]{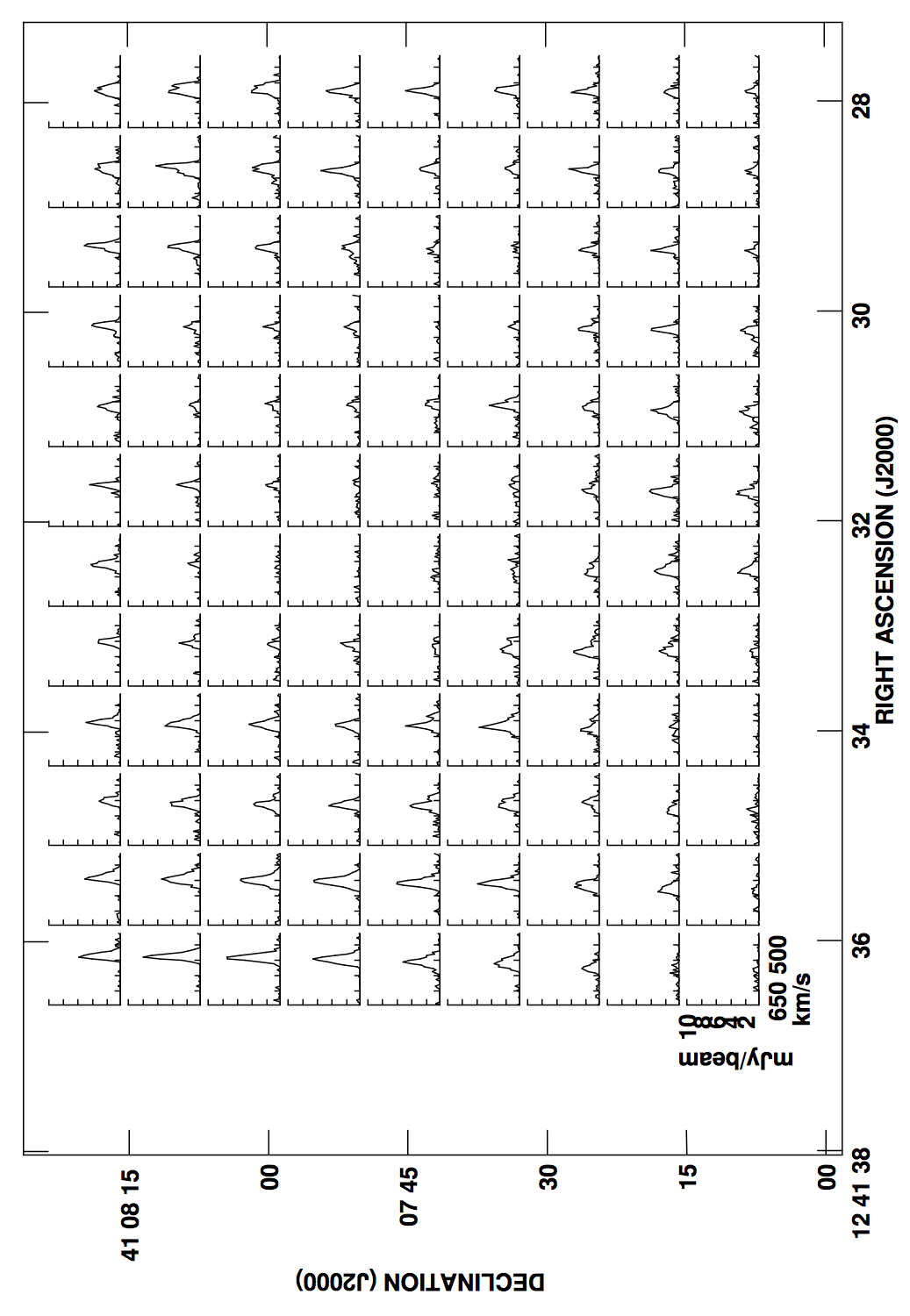}
\caption{The velocity profiles of the region surrounding holes 1, 2 3, 5 and 6, as shown in Figure 8. Each box represents the H~I profile for a region equivalent to the beam. A flux cut off of 2.00$\times$10$^{-4}$Jy B$^{-1}$ was applied to the displayed line profiles of the intermediately weighted data cube (Robust 0). The profiles seem devoid of double peaks and significant broadening, arguing against the expansion of the holes.}
\end{figure}

\clearpage

\begin{figure}
\centering
\includegraphics[width=.9\linewidth, angle=270]{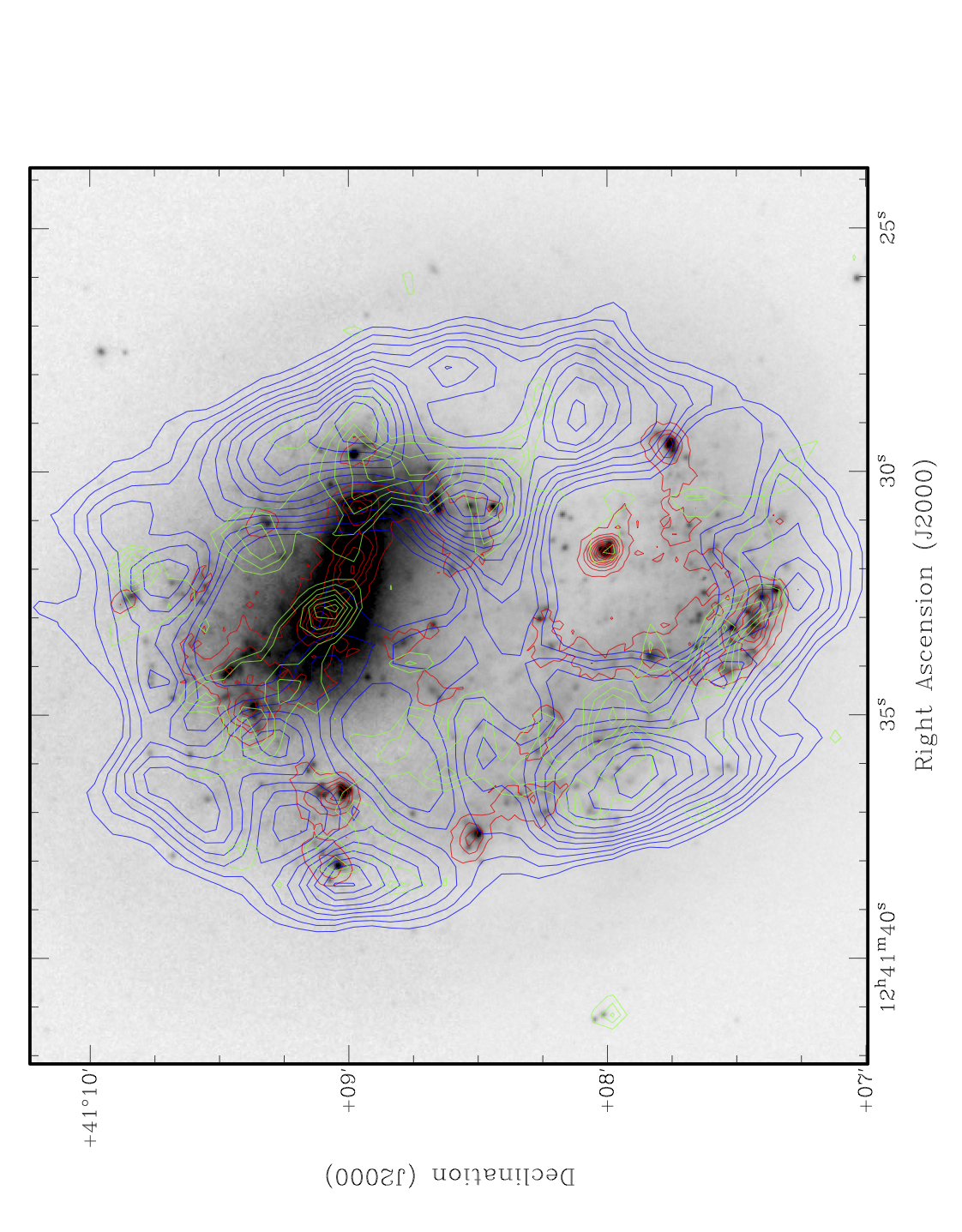}
\caption{\footnotesize SDSS optical image of NGC 4618 overlaid with NUV (red), robust 5 H~I column density (blue) and 1.4 GHz continuum (green) contours. The H~I contours correspond to every 5\% increase in density ranging from 9.63 M$_{\odot}$pc$^{-2}$ to 21.4 M$_{\odot}$pc$^{-2}$.  UV contours represent increases of 3.9$\times$10$^{-2}$ $\mu$Jy, or 10\% increments in intensity. The continuum contours range from 72 to 131 mJy B$^{-1}$ in increments of 17 mJy B$^{-1}$. These fluxes correspond to a range in star formation rate from 2.39$\times$10$^{-5}$ to 4.35$\times$10$^{-5}$ M$_{\odot}$yr$^{-1}$kpc$^{-2}$ with each contour level representing a 5.64$\times$10$^{-6}$ M$_{\odot}$yr$^{-1}$kpc$^{-2}$ change. There exists an anti-correlation between the location of the H~I holes and peaks in the UV. The strong stellar bar is shown to also have extensive UV and 1.4 GHz emission associated with it.}
\end{figure}

\begin{figure}
\centering
\includegraphics[width=.9\linewidth]{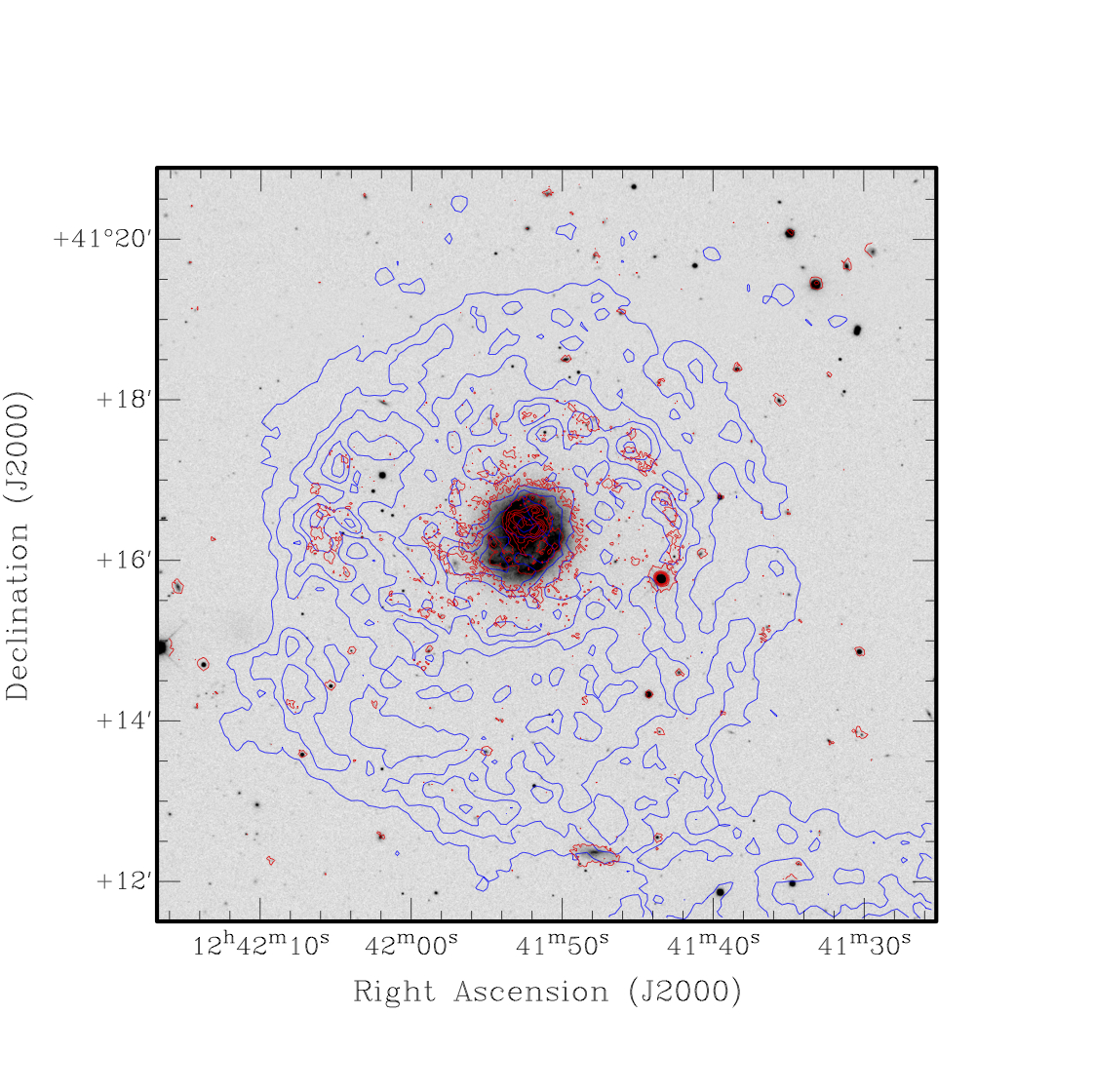}
\caption{\footnotesize SDSS optical r-band image of NGC 4625 with
\textit{GALEX} NUV (red) and H~I column density (blue). NUV contours
correspond to increments of .048 $\mu$Jy spanning
9.54$\times$10$^{-3}$ to 0.248 $\mu$Jy. H~I contours map densities of
1.1 to 6.42 M$_{\odot}$pc$^{-2}$ with increments of 1.1
M$_{\odot}$pc$^{-2}$.  It is worth noting a bright area of UV emission
(12$^{h}$41$^{m}$47$^{s}$, +41$^{\circ}$12'22''.2) south of NGC 4625
that is measured as having too low a H~I density for star
formation. Upon further investigation, it was found that the UV source
originates from galaxy J124147.76+411222.2 at a redshift of z=0.0247
as identified in the Sloan Digital Sky Survey archive. Radio continuum
emission from this galaxy could be seen in our data cube before the
continuum was subtracted in the cleaning process. }
\end{figure}

\begin{figure}
\includegraphics[width=1.0\textwidth]{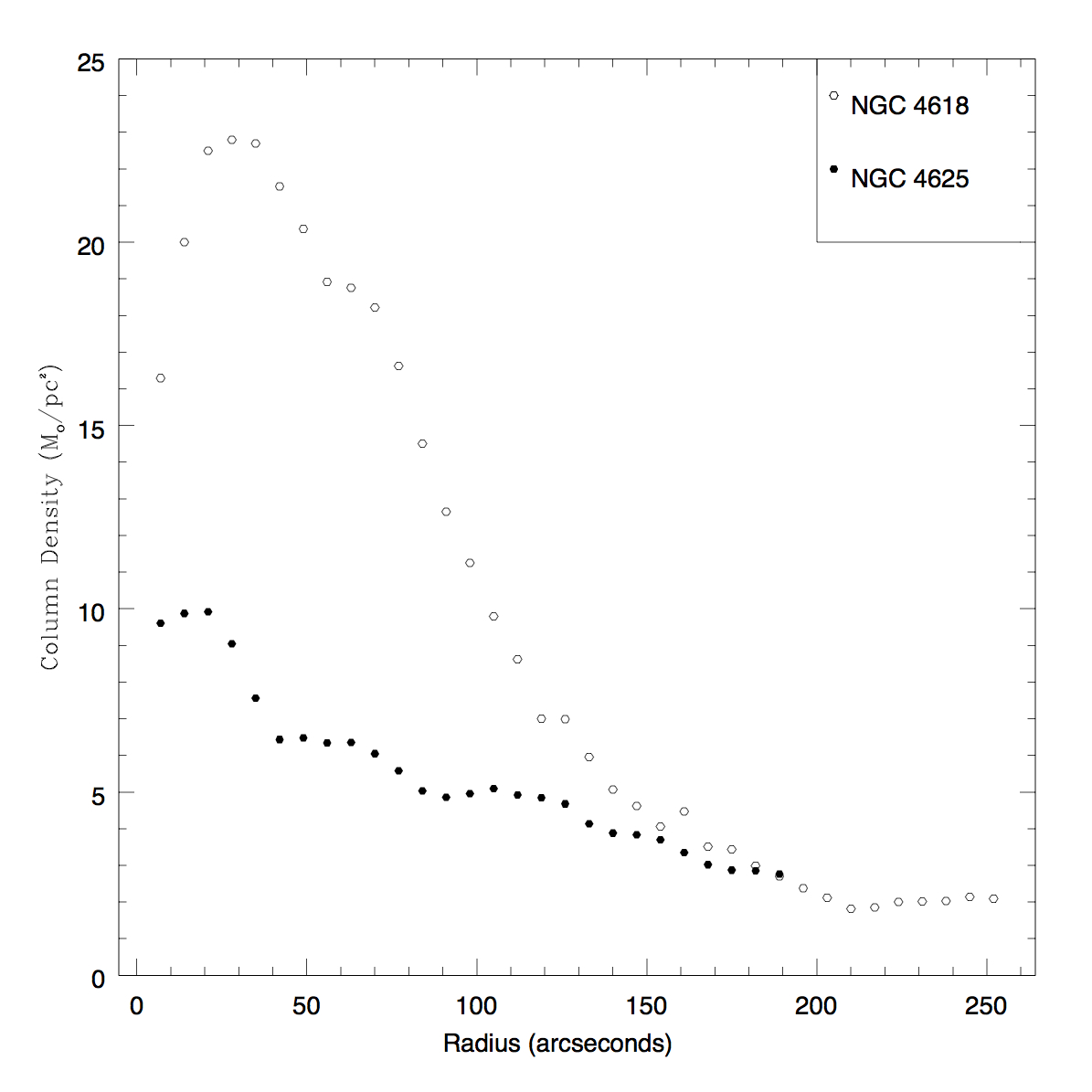}
\caption{\footnotesize Radial plots of H~I surface densities for both NGC 4618 and NGC 4625. Analysis is carried out to a radius equivalent to that of the rotation curve analysis with 7$^{\prime\prime}$ increments. The peak densities in both galaxies occurs within the core of each disk.}
\end{figure}

\clearpage

\begin{landscape}
\begin{figure}[ht]
\begin{minipage}[t]{.8\linewidth}
\centering
\includegraphics[width=.9\linewidth, angle=270]{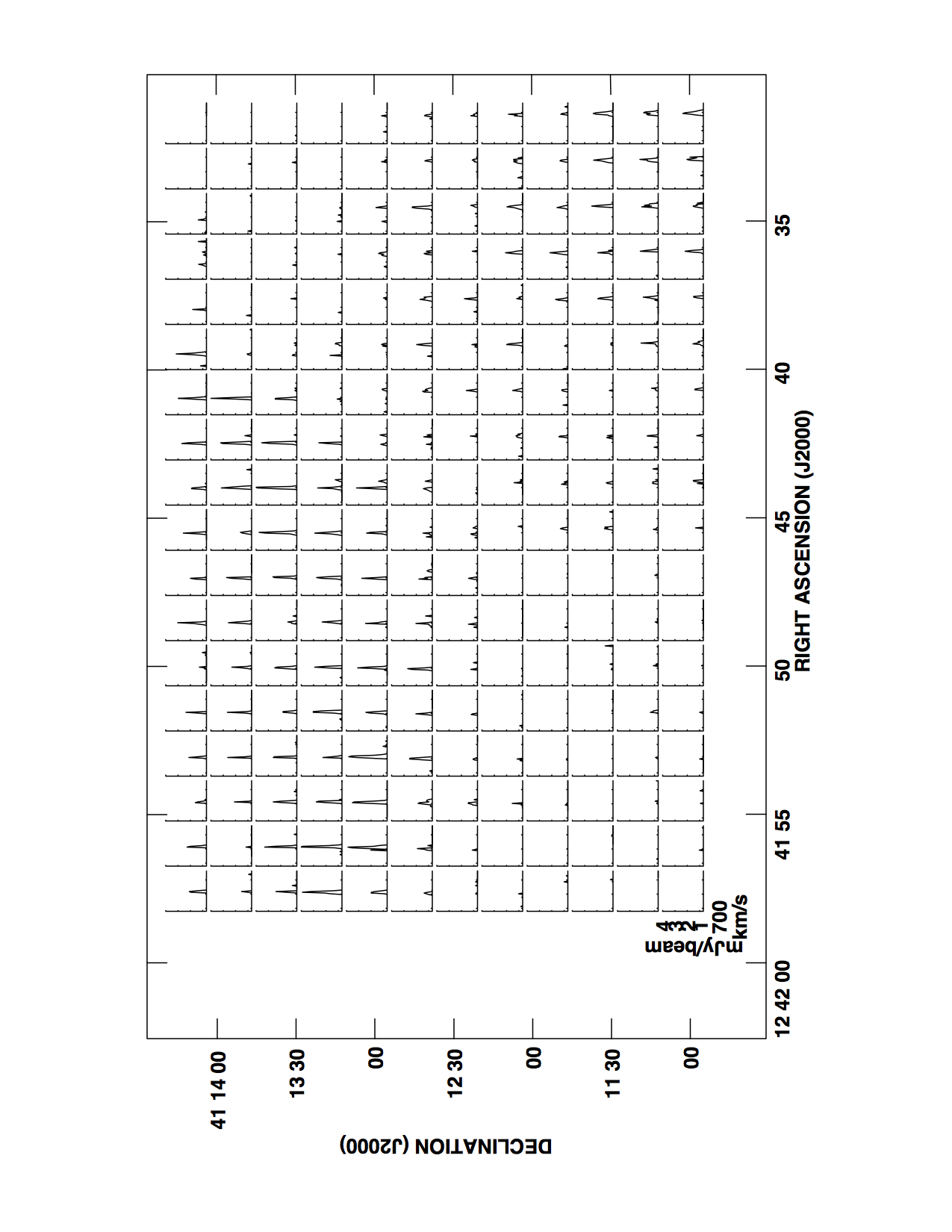}
\end{minipage}
\caption[width=1.00\linewidth]{\small Line profiles of the H~I gas corresponding to
the area that is visually suggestive of an interaction.  Each box
shows the line profile for the gas in an area equivalent to one beam
on the galaxy.  The southwestern part of NGC 4625 appears in the upper left of the image, while the northeastern part of NGC 4618 appears in the lower right.  The region of apparent overlap is in the middle of the image.  This is the same region for which we see very high velocity dispersions in Figure 4.  At first glance this would be
suggestive of an interaction between the two galaxies.  This image,
however, shows that the apparently high velocity dispersions arise
from the fact that there are two distinct kinematic components
contributing to the overlap. Thus, the existence of double peaks in
the RA range of 12$^{h}$41$^{m}$40--47$^{s}$ argues against an interaction.}
\end{figure}
\end{landscape}

\end{document}